\documentclass[twocolumn,aps,pra,reprint,groupedaddress,floatfix]{revtex4-2} 
\usepackage{graphicx}
\usepackage{amsmath,amsfonts,amssymb,physics}
\usepackage[bbgreekl]{mathbbol}
\usepackage{bm}
\newcommand{\bmmcS}{\bm{\mathcal{S} }}
\usepackage{xcolor}
\usepackage[draft,inline,nomargin]{fixme} \fxsetup{theme=color}
\FXRegisterAuthor{cp}{acp}{\color{blue}CP}
\FXRegisterAuthor{ku}{aku}{\color{teal}KU}
\FXRegisterAuthor{im}{aim}{\color{red}IM}
\FXRegisterAuthor{cc}{acc}{\color{brown}CC}
\FXRegisterAuthor{vr}{avr}{\color{olive}VR}
\usepackage[mathlines]{lineno}  
\graphicspath{{images/}}
\usepackage{tikz}
\usetikzlibrary{decorations.pathmorphing}
\usetikzlibrary{arrows.meta}
\usetikzlibrary{calc}
\tikzset{
  arrow/.style={thick, ->, >=Stealth},nodecircle/.style={circle,draw=none,minimum width=1.0cm,minimum height=1.0cm,inner sep=0pt,font=\large,align=center}}
\usepackage{stmaryrd} 
\def\>{\rangle}
\def\<{\langle}
\def\ii{{\rm i}}
\def\dd{{\rm d}}
\def\abs#1{{\scriptstyle|}#1{\scriptstyle|}}
\def\Lie#1{\hbox{\sf #1}}

\def\U#1{\Lie{U($#1$)}}
\def\onehalf{{\textstyle \frac12}}

\def\ket#1{| #1 \rangle}
\def\bra#1{{\langle #1 |}}
\newcommand{\bracket}[2]{\langle #1 | #2 \rangle}
\newcommand{\mcC}{\mathfrak{C}}

\newcommand{\rmt}{\mathrm{ts}}
\newcommand{\avs}{\mathrm{as}}
\newcommand{\eff}{\mathrm{eff}}
\renewcommand{\tr}{\text{Tr}}
\newcommand{\bq}{\begin{equation}}
\newcommand{\eq}{\end{equation}}
\newcommand{\bqa}{\begin{eqnarray}}
\newcommand{\eqa}{\end{eqnarray}}
\newcommand{\bl}{\begin{align}}
\newcommand{\el}{\end{align}}
\newcommand{\ba}{\begin{array}}
\newcommand{\ea}{\end{array}}
\newcommand{\bc}{\begin{cases}}
\newcommand{\ec}{\end{cases}}
\newcommand{\bpm}{\begin{pmatrix}}
\newcommand{\epm}{\end{pmatrix}}
\newcommand{\Eref}[1]{Eq.~(\ref{#1})} 
\newcommand{\Sref}[1]{Sec.~\ref{#1}}
\newcommand{\Fref}[1]{Fig.~\ref{#1}}  

\newcommand{\Aref}[1]{Appendix~\ref{#1}}
\newcommand{\rts}{r_\mathrm{ts}}
\newcommand{\vecrts}{\vb{r}_\mathrm{ts}}

\newcommand{\ie}{\emph{i.e.}}
\newcommand{\eg}{\emph{e.g.}}
\newcommand{\etc}{\emph{etc.}}
\begin{document} 

\title{Detecting quantum many-body states with imperfect measuring devices}
\author{Kenan Uriostegui${}^{1,2}$}
\email{rkuu@icf.unam.mx}
\author{Carlos Pineda${}^{1,3}$}
\email{carlospgmat03@gmail.com}
\affiliation{${}^{1}$Instituto de Física, Universidad Nacional Autónoma de México, Ciudad de México 04510, México\\
${}^{2}$Instituto de Ciencias Físicas, Universidad Nacional Autónoma de México, Cuernavaca, Morelos 62210, México\\
${}^{3}$Vienna Center for Quantum Science and Technology, Atominstitut, TU Wien, 1020 Vienna, Austria
}
\author{C.{} Chryssomalakos${}^{4}$}
\email{chryss@nucleares.unam.mx}
\author{V.{} Rasc\'on Barajas${}^{4}$}
\email{valentina.rascon@correo.nucleares.unam.mx}
\author{I.{} V\'azquez Mota${}^{4}$}
\email{igor.vazquez@correo.nucleares.unam.mx}
\affiliation{${}^{4}$Instituto de Ciencias Nucleares, Universidad Nacional Autónoma de México, Ciudad de México 04510, México}
\date{\today}

\begin{abstract}
We study a coarse-graining map arising from incomplete and imperfect addressing of particles in a multipartite quantum system. In its simplest form, corresponding to a two-qubit state, the resulting channel produces a convex mixture of the two partial traces. We derive the probability density of obtaining a given coarse-grained state, using geometric arguments for two qubits coarse-grained to one, and random-matrix methods for larger systems. As the number of qubits increases, the probability density sharply concentrates around the maximally mixed state, making nearly pure coarse-grained states increasingly unlikely. For two qubits, we also compute the inverse state needed to characterize the effective dynamics under coarse-graining and find that the average preimage of the maximally mixed state contains a finite singlet component. Finally, we validate the analytical predictions by inferring the underlying probabilities from Monte-Carlo-generated coarse-grained statistics.
\end{abstract}
\maketitle
\section{Introduction \label{sec:Introduction}} 

Our knowledge of the natural laws governing a physical system fundamentally
relies on our measurement capabilities. Bridging the gap between a system's
intrinsic nature and our understanding of it depends crucially on both the
measurement devices we employ and the theoretical frameworks we use to
interpret the obtained information. In this regard, Einstein famously remarked
to Heisenberg during an informal conversation: ``It is the theory which decides
what we can observe. You must appreciate that observation is a very complicated
process'' \cite{Heisenberg1971}. Years later, Heisenberg echoed this insight,
stating: ''We have to remember that what we observe is not nature herself, but
nature exposed to our method of questioning'' \cite{Heisenberg1958}.
Thus, recognizing the limits of our measurements is crucial to understanding what the quantum theory can actually tell us.

These considerations motivate us to address specific problems in the detection of quantum many-body states. Detecting and
characterizing quantum states is essential for the validation, control, and
optimization of quantum devices, becoming increasingly critical in modern
technological advancements \cite{Eisert2020, Altman2021, Pelucchi2022}.
However, reconstructing states in quantum many-body systems presents significant challenges. One of the main issues stems from intrinsic measurement uncertainties.
Measurement devices are susceptible to
systematic errors even under optimal experimental conditions, significantly
affecting the accurate characterization of states in complex many-body quantum
systems \cite{Rosset2012}. Therefore, such measurement imperfections cannot be ignored and must be carefully managed.

Several proposals have been put forward to address the issue of imperfect
measurements, both from a fundamental perspective and in terms of their impact on applications.
Among them are approaches based on generalized measurements,
characterized by parameters such as bias and sharpness, which enable the
modeling of various effects contributing to measurement non-ideality
\cite{Naikoo2021}. Compatibility criteria between fuzzy measurements have also
been investigated \cite{Wu2011}. More recently, methods involving neural
networks—trained to recognize quantum states in a manner analogous to blurred
image recognition—have been proposed, along with active learning techniques
designed to optimize data collection during the characterization process
\citep{Zhu2023,Gao2022,Lange2023}. While these strategies have enabled
significant progress in state reconstruction and the identification of
noise-resilient patterns, they face notable limitations. These include their
reliance on system-specific training data, the challenges associated with
extending them in a controlled way to multipartite systems, and the lack of a
systematic framework that directly links model parameters to the operational
features of the measurement devices.

A systematic framework based on a coarse-graining map for studying quantum
many-body systems under imperfect measurements was proposed 
in \cite{Pineda2021}. This map, denoted by $\mathcal{C}$, applies to systems
composed of $N$ particles and is constructed from a fuzzy measurement followed by
a reduction in the system's degrees of freedom. The resulting procedure defines
a quantum channel designed to model experimental errors stemming from
particle-indexing ambiguities and from the intrinsic limitations in the ability to precisely identify the number of particles composing the system. 
The map $\mathcal{C}$ complies with
Uhlmann's theorem and predicts a doubly exponential reduction in the volume of
distinguishable states as the number of particles increases, which aligns with
the challenges faced in quantum tomography. 

Coarse-graining maps use well-defined rules to produce a simpler description of a physical system from a more complete one, accounting for the different ways in which information can be lost  \cite{Busch1993}. This makes it possible to describe not only the system at a given moment, but also how its effective dynamics relate to the fine dynamics of the full description.
If the initial state is obtained through measurements modeled by a coarse-graining map, it is called a granular state. After the system evolves, measuring it again gives its granular or effective dynamics. This dynamics is related to the fine dynamics —which link fine states in the complete description— through a map that acts as a bridge between the two \cite{Quadt1994,Duarte2017}.
Therefore, knowing this map is essential for fully addressing the problem of imperfect measurements.

In this work, we adopt the $\mathcal{C}$ map to analyze the many-body
(fine-grained) states that give rise to the detection of coarse-grained states
under imperfect measurements. We systematically characterize the set of
fine-grained states that are mapped to a small neighborhood of a given monopartite state. This leads to
the definition of the preimage set, whose volume determines the probability
density function (PDF) associated with a target single-qubit state obtained
from an $N$-qubit system. The PDF serves as a fundamental tool for analyzing
how observable properties, quantum coherence, and correlations are affected by
the coarse-graining process. In particular, it allows us to identify quantum
features that remain robust under imperfect measurements described by this map.

Building on the characterization of preimage sets, we define an ``inverse'' map by establishing a rule to assign a preimage state to each target state. By definition, the coarse-graining (CG) map is not 1-to-1 and thus cannot be strictly inverted. Nevertheless, physically motivated criteria can be introduced to select a specific fine-grained preimage for each target state. For example, in \cite{Castillo2024} the authors propose choosing the fine-grained state of maximum entropy compatible with the given target state. Motivated by quantum state tomography procedures—where reconstructed states naturally correspond to averages over multiple experimental realizations—we assign, as the preimage of a target state, the average of all fine-grained states that the coarse-graining map $\mathcal{C}$ sends within a distance $\epsilon$ of the target state. We refer to this rule as the {\it average assignment} and denote it by $\overline{\mathcal{C}^{-1}}$. We then compute and analyze the average state in the case where the system is coarse-grained from two particles to one.

We focus on pure states of systems composed of $N$ qubits, analyzing the
scenario in which imperfections in the measurement device lead to the
tomographic reconstruction of coarse-grained states describing a single qubit.
Within this framework, we address two fundamental questions: What is the most
representative fine-grained state corresponding to the observed coarse-grained
state, given that it was obtained through a device subject to indexing errors
and limited in its ability to resolve all degrees of freedom of the system? And
how do such imperfections impact our knowledge of the quantum properties of
the original system? 
To address these questions, we employ
geometric methods
and random matrix theory (RMT) to compute the volume of the preimage set when
the coarse-graining map $\mathcal{C}$ reduces the system from $N$ qubits to
one, and we compute the average state when $\mathcal{C}$
reduces a two-qubit system to a single qubit. 

The article is organized as follows. \Sref{sec:cgm} reviews the structure and main
properties of the coarse-graining map $\mathcal{C}$. In \Sref{sec:Probability-target-state}, we compute
and analyze the PDF for a two-qubit system reduced to a single qubit, using
geometrically motivated parameterizations and analytical methods from RMT. These RMT-based
techniques are then generalized to obtain the PDF for the more general case of
an $N$-qubit system reduced to a single qubit. \Sref{sec:average-state} focuses on the
computation and characterization of the average state in the two-to-one qubit
case, emphasizing its symmetries and structural features. In \Sref{sec:statistics}, we
compare our theoretical predictions with numerical results obtained via Monte
Carlo simulations and propose practical strategies for identifying
coarse-graining effects in experimental scenarios. Finally, \Sref{sec:Conclusions} presents
our conclusions and outlines possible directions for future research.
\section{Coarse-graining map\label{sec:cgm}} 

In this section, we introduce the coarse-graining (CG) map, which forms the
basis of our study \cite{Pineda2021}. We define how the $\mathcal{C}$ map is
constructed and describe its role in our approach. We also point out a symmetry
of the map that will be important in later sections.

The imperfect detections considered in this study arise from two main
limitations of the measurement device. First, it is not possible to identify
each particle in the system with complete accuracy. For example, in
two-dimensional ultracold atomic systems, atoms are detected by fluorescence
imaging with a high-resolution microscope. An algorithm determines the presence
of an atom based on its brightness and proximity to predefined grid points on
the image \cite{Sherson2010,Yang2021,Kwon2022}. Because the grid is discrete
and has finite resolution—typically on the order of hundreds of
nanometers—there is an inherent indexing error in identifying particles. These
errors reduce the precision of the measurement and are referred to as fuzzy
measurements (FM).

The simplest case is where we deal with a bipartite system. Taking into account that the measurement device has some probability $p$ of mistaking the two particles that determine the state $\varrho$. This situation is described as a fuzzy measurement of the form
\bq
\mathcal{F}[\varrho] = (1-p)\varrho + pP[\varrho],
\eq
where the brackets denote the action of the permutation operator on the density
matrix, $P[\varrho]=P\varrho P^{\dagger}$. Then, in the general situation where we have an $N$-partite system, our imperfect measurement device can detect any possible permutation of the $N$ particles with some probability $p_{P}\geq 0$, so the FM takes the form 
\bq
    \mathcal{F}[\varrho] = \sum_{P\in \mathcal{S}_{N}}p_{P}P[\varrho],\qquad \sum_{P\in\mathcal{S}_{N}}p_{P} =1, \label{eq:fuzzymeasurement}
\eq
where $P$ belongs to a subset of the symmetric group $\mathcal{S}_{N}$ of
$N$ particles.

The second type of imperfection we consider arises from the measurement device’s inability to resolve the fine details of the whole system, leading to errors in identifying the number of particles composing the system.
As a consequence, groups of particles are collectively represented within effective states corresponding to a reduced number of particles.
This coarse-grained approach to observing the system gives rise to the term coarse-grained measurement.
Both types of errors are deeply related, as illustrated by the example of a two-dimensional quantum system, where two or more atoms can appear indistinguishable behind a single bright spot. Their presence implies that the way particles are grouped collectively depends on $p_P$, meaning that such groups are not sharply determined.
It is worth noting, however, that coarse-graining of the system can arise either from the limited sensitivity of the measurement device—as in the cited example—or because only a few degrees of freedom are relevant to the specific property under study. 

Suppose a detector is capable of identifying only $n$ particles among the $N$
comprising a system. Since it cannot distinguish the remaining $m=N-n$
particles, in each measurement it randomly selects a specific subset of $n$
particles and disregards the information about the others. This procedure is
mathematically equivalent to performing a partial trace over a fuzzy
measurement of the entire system, thus defining the coarse-graining (CG) map
\begin{gather}
    \mathcal{C} : \mathcal{B}(\mathsf{H}_{d}^{\otimes N}) \longrightarrow \mathcal{B}(\mathsf{H}_{d}^{\otimes n}) \\[5pt]
    \mathcal{C}[\varrho] = \tr_{\bar{\mathsf{n} }} \mathcal{F}[\varrho] \label{eq:coarse-graining}
\end{gather}
where $d$ is the number of levels for each particle in the system,
$\bar{\mathsf{n}}$ denotes the complementary subset to the identified $n$
particles (such that $\abs{\bar{\mathsf{n}} }=m$), and
$\mathcal{B}(\mathsf{H})$ represents the space of bounded linear operators
acting on $\mathsf{H}$.

We consider a system of qubits ($d=2$) measured by a device that can detect only one particle ($n=1$). Without loss of generality, we label the detectable particle as particle 1. As a result, the measurement can only exchange this particle with any of the others, and only the swap operations $S_{1,i}$ are relevant, since all other permutations vanish once the non-measured particles are traced out.
For simplicity, we restrict the discussion to pure states $\varrho=\ket{\psi}\bra{\psi}$ with $\ket{\psi}\in\mathsf{H}_2^{\otimes N}$. Under these assumptions, the coarse-graining map becomes
\begin{equation} 
    \mathcal{C}[\ket{\psi}\bra{\psi}] = \tr_{\bar{\mathsf{1} }} 
         \sum_{i=1}^{N}p_{1i}S_{1,i}[\ket{\psi}\bra{\psi}]. 
	 \label{eq:cgmbruto}
\end{equation} 
For each swap, the particle brought to the detector determines the outcome of the partial trace, which becomes its reduced density matrix.
Thus, we can express the map (\ref{eq:cgmbruto}) in the simplified form
\bq 
    \mathcal{C}\left[\ket{\psi}\bra{\psi} \right] = \sum_{i=1}^{N} p_{i}\rho_{i}, \label{eq:def-CGM}
\eq
with
\bq 
    \rho_{i} = \tr_{\bar{\mathsf{i} }}\ket{\psi}\bra{\psi}\qquad \sum_{i=1}^{N} p_{i} = 1, \label{eq:complementoCGM}
\eq
where $\rho_{i}$ denotes the reduced density matrix associated with the $i$-th particle, and $p_{i}$ is the probability of detecting that particle.

The form given by (\ref{eq:def-CGM})-(\ref{eq:complementoCGM}) highlights a
symmetry property of the CG map. Specifically, the compatibility of the partial
trace with unitary transformations \cite{Ibrahim2015}, expressed as 
\bq
    U \ket{\psi_{i}}\bra{\psi_{i}} U^{\dagger} = \tr_{\bar{\mathsf{i}} } U^{\otimes N} \ket{\psi}\bra{\psi}U^{\dagger\otimes N},
\eq
where $\ket{\psi} = \bigotimes_{i=1}^{N} \ket{\psi_{i}}$, directly leads to the following unitary covariance relation for the map,
\bq 
	\mathcal{C}\left[U^{\otimes N}\ket{\psi}\bra{\psi}U^{\dagger\otimes N}\right]
	= U \mathcal{C}[\ket{\psi}\bra{\psi}] U^{\dagger}. \label{eq:simetria-map}
\eq
where $U$ is a unitary operator in $\mathcal{B}(\mathsf{H}_{2})$. Later on, we will leverage this property to reveal characteristics of certain average states.

\section{Probability of a target state \label{sec:Probability-target-state}} 
Consider the physical situation in which measurements are performed on a system of $N$ qubits using imperfect devices that can only resolve measurements on one of the particles, described by a channel like in~\Eref{eq:def-CGM}. Since it is not possible to access the full description of the system, we assume that an ensemble of random $N$-partite states is available.

The question then arises, what is the probability that the observed state falls within a small volume $V_\epsilon$, surrounding a given target state $\rho=(I+\vecrts \cdot \vb*\sigma)/2$? Here, $\vb*\sigma$ is a vector whose entries are the Pauli matrices, $\vecrts$ is a vector in the Bloch ball $B$ representing $\rho$, and the volume $V_\epsilon$ corresponds to the region defined by the usual measurement uncertainty bars --- the region within which the observed state can be guaranteed to lie.

To answer this question, it is necessary to establish the probability distribution followed by the fine-grained states. In this section, we address the $N=2$ case in detail, considering both the situation where the fine-grained states are uniformly distributed with respect to the Fubini–Study measure on the full space, and the situation where this distribution is restricted to the space of separable states. For the $N$-to-$1$ case, only the first scenario will be analyzed. 

To this end, we describe the probability measure in the pure state manifold. We start with the set of $N$-partite pure states which, after fixing the normalization and discarding the overall phase, correspond to the complex projective space $\mathbb{C}P^D \equiv \mathbb{P}$, with $D=2^N-1$, whose points represent $2^N \times 2^N$ pure state density matrices. An immediate result obtained by analyzing the r.h.s.{} of~\Eref{eq:def-CGM} for separable and maximally entangled states, is that the image of the map $\mathcal{C}$ covers the entire Bloch ball $B$ for any choice of the probabilities $p_i$.

The projective space $\mathbb{P}$ is equipped with a metric inherited from the structure of the Hilbert space $\mathcal{H}\simeq \mathbb{C}^{2^N}$, known as the Fubini–Study (FS) metric~\cite{Bengtsson2008}. This metric is characterized by being the only one --- up to a multiplicative constant --- in $\mathbb{P}$ that remains invariant under the action of the unitary group $U(2^N)$. We denote by $\dd\mu$ the normalized volume measure associated with the FS metric, which can also be described as the measure induced on $\mathbb{P}$ by the Haar measure on the group $U(2^N)$.

We denote by $\Omega_{\vb{R}}^N$ the preimage of $\vb{R} \in B$ under $\mathcal{C}$,
\begin{equation}
	\Omega_{\vb{R}}^N =
	 \qty{\varrho: \mathcal{C}\qty[\varrho]=\frac{1}{2}\qty(I+\vb{R}\cdot\vb*\sigma)}\, ,
	 \label{OmegaR}
\end{equation}
and by $\Omega_\epsilon^N$ the preimage of $V_\epsilon$,
\begin{equation}
	\Omega_\epsilon^N =
	 \qty{ \bigcup_{\vb{R}} \Omega_{\vb{R}}^N : \vb{R}\in V_\epsilon}
	 \, .
	 \label{OmegaVe}
\end{equation}
Since $\int_{\mathbb{P}}\dd\mu=1$, the volume measure $\dd\mu$ defines the probability measure invariant under unitary transformations on $\mathbb{P}$. Then, the probability of observing a coarse-grained state within $V_\epsilon$ is obtained by computing the volume of the preimage set $\Omega_\epsilon^N$ with respect to $\dd\mu$:
\begin{equation}
	V(\Omega_\epsilon^N) = \int_{\Omega_\epsilon^N}\dd\mu 
	\, .
	\label{eq:Def-volumen}
\end{equation}

In what follows, we also consider the case of fine-grained product states, in order to highlight the role of entanglement. We therefore define the sets
\begin{align}
	\Omega_{\vb{R}}^{N\otimes}
	&=
	\qty{ \varrho =\textstyle\bigotimes\limits_{i} \rho_i : 
	       \mathcal{C}\qty[\varrho]=\frac{1}{2}\qty(I+\vb{R}\cdot\vb*\sigma)}
	\, ,
	\label{OmegaVR}
	\\
	\Omega_\epsilon^{N\otimes}
	&=
	\qty{ \bigcup_{\vb{R}} \Omega_{\vb{R}}^{N\otimes} : \vb{R}\in V_\epsilon}
	\, ,
	\label{OmegaVeot}
\end{align}
and use $\dd\mu^{\otimes}$ to denote the FS measure restricted to the subspace of product states in $\mathbb{P}$. Other variants of the above problem can also be considered; for example, when dealing with states that possess particular symmetries, the integration in \Eref{eq:Def-volumen} can be restricted to the appropriate subspace.

In the rest of this section, we first carry out the integration in~(\ref{eq:Def-volumen}) for the case $N=2$. We give an appealing geometric description that allow us to visualize the sets $\Omega_{\epsilon}^2$ and $\Omega_{\epsilon}^{2 \otimes}$ and compute the corresponding integrals. To keep the notation simple, we omit the superscript 2 when referring to this case.

For $N>2$, the geometric approach becomes unmanageable, and we resort to random matrix techniques to obtain an exact result. As preparation, we first study the case $N=2$ and then address the general case. We conclude by discussing several physical consequences of the results obtained.

\subsection{Bipartite system} 
\subsubsection{Parametrization of pure two-qubit states} 

We now turn to a convenient parametrization of two-qubit pure states. This parametrization is defined in terms of (i) the coordinates of the reduced density matrices of each qubit, and (ii) two nonlocal parameters: one quantifying the degree of entanglement between the qubits, and another representing a relative phase, which will not play a role in our present
analysis.


We now work through the technical details of such parametrization.
The Schmidt decomposition of a 2-qubit state involves a rotation of a three-dimensional coordinate system associated with each particle in such a way that an axis of each coordinate system lines up with the Bloch vector of the corresponding particle while the correlation tensor between the two qubits, $C_{kl} = \tr\left((\sigma_{k}\otimes \sigma_{l})\,\ket{\psi}\bra{\psi}\right)$, 
becomes diagonal \cite{Ekert1995, Aravind1996}.
As a result, a general normalized 2-qubit pure state $\ket{\psi}$ can be written in the form
\begin{equation}
\label{psiSdec}
\ket{\psi}
=
\sqrt{\lambda} \ket{e_1, f_1} +\sqrt{1-\lambda} \ket{e_2, f_2}
\, ,
\end{equation}
where  $\lambda$ is real and non-negative, and $\{\ket{e_i}\}$, $\{\ket{f_i}\}$
are particular orthonormal bases, in the single-qubit Hilbert space, that
depend on $\ket{\psi}$. Taking
$\{\ket{e_1},\ket{e_2}\}=\{\ket{\vb{n}_1},\ket{{}-\vb{n}_1}\}$,
$\{\ket{f_1},\ket{f_2}\}=\{\ket{\vb{n}_2},e^{i \gamma}\ket{{}-\vb{n}_2}\}$,
where $\ket{\vb{n}_i}=\cos \frac{\theta_i}{2} e^{-i\frac{\phi_i}{2}}\ket{0}
+\sin\frac{\theta_i}{2} e^{i\frac{\phi_i}{2}}\ket{1}$, and putting
$\sqrt{\lambda}=\cos\frac{\eta}{2}$, $0 \leq \eta \leq \pi$, we get
a 
\begin{align}
	\ket{\psi} 
	&= 
	\cos\frac{\eta}{2}\ket{\vb{n}_1,\vb{n}_2}+\sin\frac{\eta}{2}e^{\ii\gamma}\ket{-\vb{n}_1,-\vb{n}_2}, \nonumber \\
	&=  
	e^{-\frac{1}{2} \ii (\phi_{1} + \phi_{2})} 
 	\big(\cos\frac{\eta}{2}\cos\frac{\theta_{1}}{2}\cos\frac{\theta_{2}}{2}
 	\nonumber 
 	\\
 	& 
 	\hspace{2.2cm}  +\, e^{\ii\gamma} \sin\frac{\eta}{2}\sin\frac{\theta_{1}}{2}
 	\sin\frac{\theta_{2}}{2}\big)\ket{0,0} \nonumber 
 	\\
 	& 
 	+\, e^{-\frac{1}{2}\ii (\phi_{1} - \phi_{2})} \big(\cos\frac{\eta}{2}
 	\cos\frac{\theta_{1}}{2}\sin\frac{\theta_{2}}{2} \nonumber 
 	\\
 	& 
 	\hspace{2.2cm} -\, e^{\ii\gamma} \sin\frac{\eta}{2}\sin\frac{\theta_{1}}{2}
 	\cos\frac{\theta_{2}}{2}\big)\ket{0,1} \nonumber 
 	\\
 	& +\, e^{\frac{1}{2}\ii (\phi_{1} - \phi_{2})} \big(\cos\frac{\eta}{2}
 	\sin\frac{\theta_{1}}{2}\cos\frac{\theta_{2}}{2} \nonumber 
 	\\
 	& 
 	\hspace{2.2cm} -\, e^{\ii\gamma} \sin\frac{\eta}{2}\cos\frac{\theta_{1}}{2}
 	\sin\frac{\theta_{2}}{2}\big)\ket{1,0} 
 	\nonumber 
 	\\
 	& 
 	+\, e^{\frac{1}{2}\ii (\phi_{1} + \phi_{2})} \big(\cos\frac{\eta}{2}
 	\sin\frac{\theta_{1}}{2}\sin\frac{\theta_{2}}{2} 
 	\label{eq:app-parametrized-state} 
 	\\
 	& 
 	\hspace{2.1cm} +\, e^{\ii\gamma} \sin\frac{\eta}{2}\cos\frac{\theta_{1}}{2}
 	\cos\frac{\theta_{2}}{2}\big)\ket{1,1}
 	\, . 
 	\nonumber
\end{align}

The reduced density matrices are now easily computed,
\bq
	\rho_{i} = 
	\frac{1}{2} \begin{pmatrix}
	1+\cos\eta\cos\theta_{i} & e^{-\ii \phi_{i}}\cos\eta\sin\theta_{i} \\
	e^{\ii \phi_{i}}\cos\eta\sin\theta_{i} & 1-\cos\eta\cos\theta_{i}
	\end{pmatrix}
	\, ,
	\label{eq:app-reduced-states} 
\eq
giving for the corresponding Bloch vectors,
\begin{equation}
	\vb{r}_{i} = \tr\left( \vb*{\sigma}\rho_{i}\right) 
	= r \left(\cos\phi_{i}\sin\theta_{i},\sin\phi_{i}\sin\theta_{i},\cos\theta_{i} \right)
	\, ,
	\label{eq:app-Bloch-vector-reduced} 
\end{equation}
the radius of which  is $r=\cos\eta$. Note that $\theta_{i}$ is the polar angle and $\phi_{i}$ is the azimuthal angle of the Bloch vector $\vb{r}_{i}$. 
$r$ and $\eta$ are related to the entanglement between the two qubits. Using the concurrence $\mcC$  to measure entanglement \cite{firstconcurrence} we have that $\mcC = \sin\eta = \sqrt{1-r^2}$. Hence we shall 
refer to $\eta$ as the concurrence angle.
Note that the overall phase factor of $\ket{\psi}$ has been fixed so that the coefficient of $\ket{\vb{n}_1,\vb{n}_2}$ be real and non-negative. 

\begin{figure} 
\centering
\begin{tabular}{ccc}
    Qubit 1 & Entanglement & Qubit 2 \\
      & hemisphere & \\
    \includegraphics[width=0.33\linewidth]{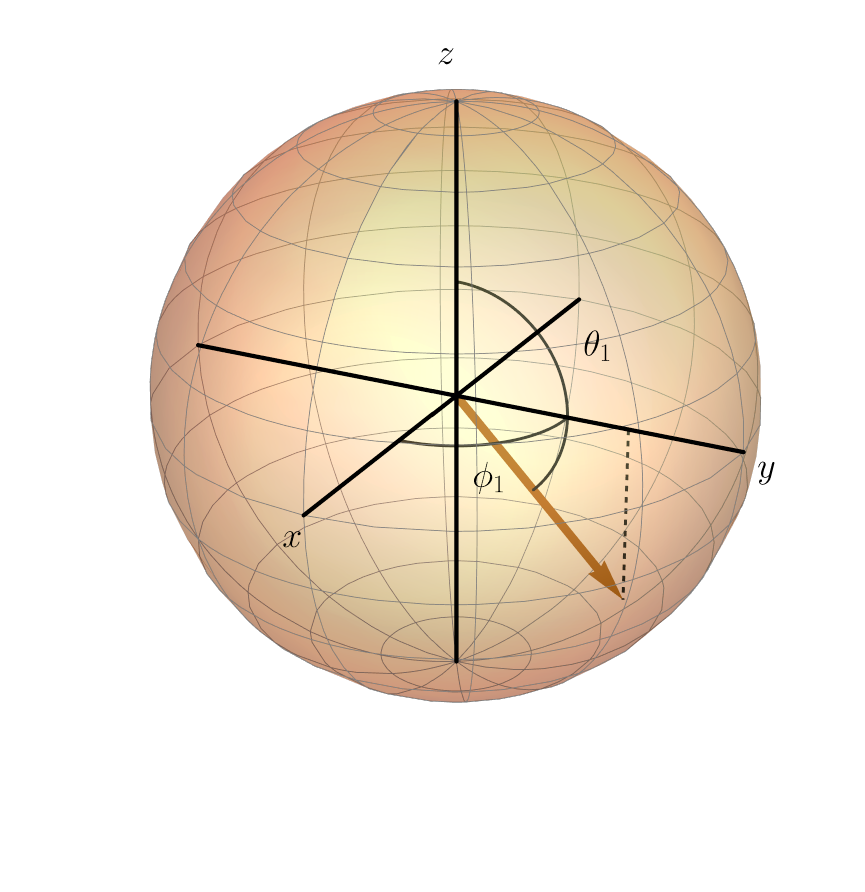} & \includegraphics[width=0.33\linewidth]{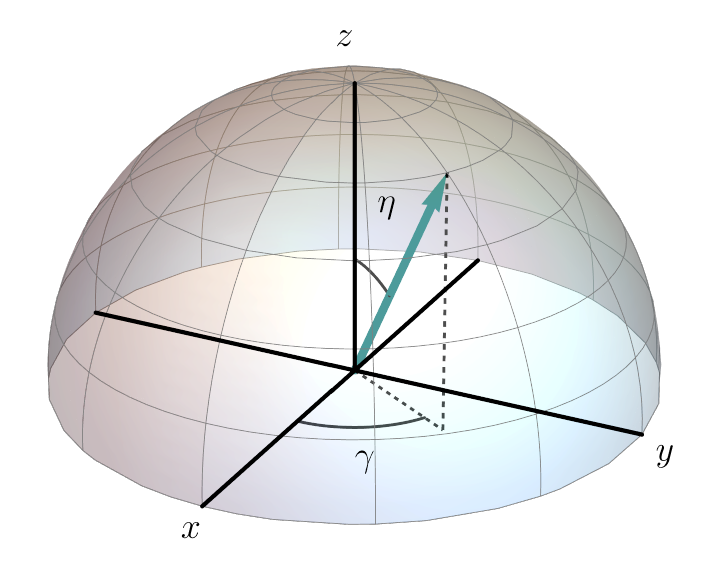} & \includegraphics[width=0.33\linewidth]{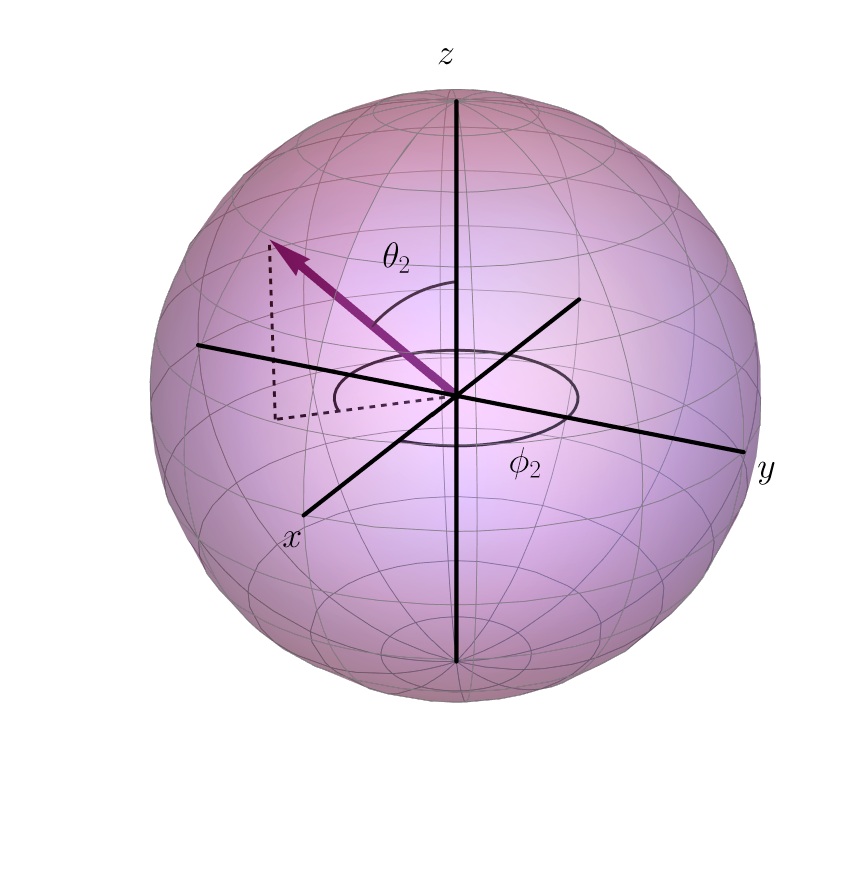} \\
\end{tabular}
\caption{
Visualization of a pure state formed by two qubits in the parameterization  
(\ref{eq:app-parametrized-state}). The state shown corresponds to the values
$\eta=\textstyle{\frac{\pi}{6}}$,$\gamma=\frac{\pi}{3}$, $\theta_{1}
=\textstyle{\frac{3\pi}{4}}$, $\phi_{1}=\frac{\pi}{2}$, $\theta_{2}=\frac{\pi}{4}$ 
and $\phi_{2}= \textstyle{\frac{7\pi}{4}}$. The orange and purple spheres correspond 
to the Bloch spheres of the reduced states of qubit 1 and 2, respectively. 
The Bloch vectors in both cases have a length $\cos\eta=\textstyle{\frac{\sqrt{3}}{2}}$. 
The gray hemisphere between the spheres is the entanglement hemisphere, it shows  
the concurrence angle $\eta$ and the relative phase $\gamma$. Note that the reduced states 
can only be pure when the entanglement arrow points to the north pole, i.e., when $\eta=0$.
\label{fig:spheres-two-qubits}}
\end{figure}

As a result, the state $\ket{\psi}$ depends on six parameters, which provide
coordinates on the corresponding projective space $\mathbb{P}$. A visualization
of these coordinates involves the Bloch spheres of each qubit in the system and
the {\it entanglement hemisphere} between them. The position of the Bloch
vectors is determined by the parameters $(\theta_{1},\phi_{1})$ and
$(\theta_{2},\phi_{2})$ and both have magnitude $\cos\eta$.  In the
entanglement hemisphere, the unit vector is oriented by the concurrence and
relative phase angles $(\eta,\gamma)$. \Fref{fig:spheres-two-qubits}
shows an example visualization of a bipartite state in this parameterization.

\subsubsection{Coarse-graining statistics via preimage volume}\label{sec:preimage_volume} 
Since the CG map in \Eref{eq:def-CGM} treats all subsystems equally, in the case $N = 2$ we assume that $p_1 < p_2$ without loss of generality. In this way, we write the coarse-grained state as $\rho = p\rho_{1} + (1-p)\rho_{2}$ with $0 \leq p \leq \onehalf$. 
In terms of Bloch vectors, this is expressed as
\begin{equation}
\vecrts = \frac{1-h}{2}\vb{r}_1 + \frac{1+h}{2}\vb{r}_2 
\, ,
\label{eq:bipartite CGM}
\end{equation}
where $\vb{r}_1$ and $\vb{r}_2$ are the Bloch vectors corresponding to the
reduced density matrices $\rho_1$ and $\rho_2$ and the symmetric parameter $h\equiv 1-2p$ is introduced. Notice that $h$
quantifies the degree of precision in the detection process. The value $h=0$
represents maximum uncertainty, where the measuring apparatus provides no information about which particle was detected, whereas $h=1$ corresponds to identification of particle $2$, occurring with unit probability.

The requirement of invariance under the action of the unitary group on $\mathbb{P}$, specifies, up to an overall scale factor, the invariant Fubini-Study metric \cite{Bengtsson2008}
\begin{multline}
\label{FSmdef}
(\dd s)^2=\bracket{\psi}{\psi}^{-2} 
\big( 
\bracket{\psi}{\psi} (\dd \bra{\psi}) (\dd \ket{\psi}) \\
-\bra{\psi}\dd \ket{\psi}(\dd \bra{\psi}) \ket{\psi}) \big)
\, ,
\end{multline}
where $\dd=dx^i \frac{\partial}{\partial x^i}$ is the exterior derivative on
$\mathbb{P}$, and $\{x^i\}$ are the coordinates
$\{r,\gamma,\theta_1,\phi_1,\theta_2,\phi_2\}$. The corresponding invariant measure is given by $\dd \mu=\mathcal{N} \sqrt{\det g} \, d^6 x$ where $g$ is the metric corresponding to the line element $(\dd s)^2 = g_{\mu\nu} \dd x^{\mu}\dd x^{\nu} $ in~(\ref{FSmdef}), and with $\mathcal{N}$ a normalization factor determined by the requirement $\int_\mathbb{P} \dd \mu =1$. Substituting $\ket{\psi}$ from \Eref{eq:app-parametrized-state} into \Eref{FSmdef} we find for the probability measure
\begin{equation}
\dd \mu= \frac{3}{2\pi}r^{2}
	\dd r\dd\gamma\dd\omega_{1}\dd\omega_{2}
	\, ,
	\label{eq:vol element}
\end{equation}
where $\dd\omega_i=\sin\theta_i \dd \theta_i \dd \phi_i/4\pi$. Restricted to product
states, the probability measure, computed similarly, takes the form 
\begin{equation}
\label{FSdmuprod}
\dd\mu^\otimes
=\dd \omega_1 \dd \omega_2
\end{equation}
\emph{i.e.}, it just factorizes into the product of the measures of two Bloch
spheres. 

\begin{figure}[htbp] 
\begin{tabular}{cc}
\includegraphics[width=0.5\linewidth]{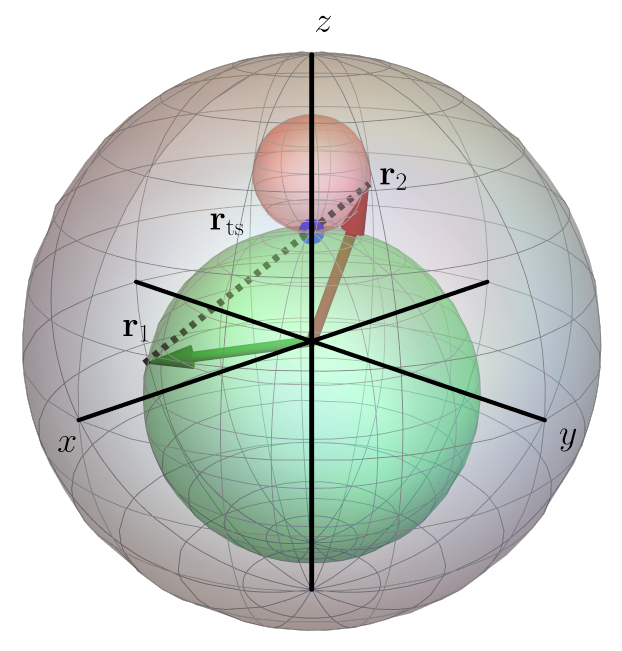} & \includegraphics[width=0.5\linewidth]{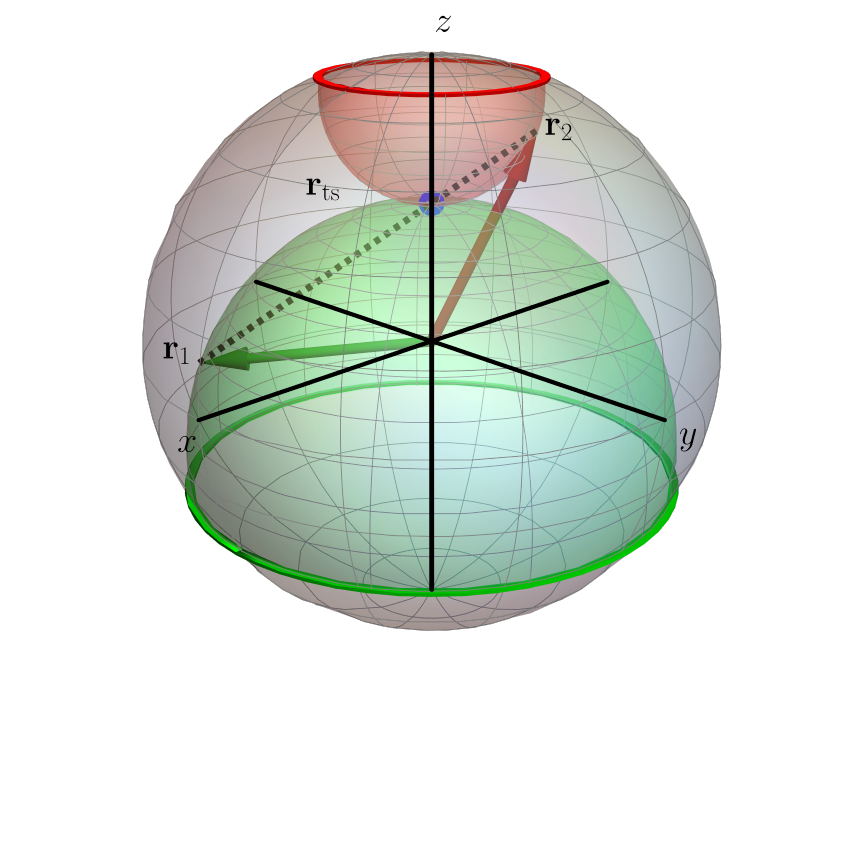}
\end{tabular}
\caption{
Green and red subspheres corresponding to the locus of states $\rho_{1}$ and
$\rho_{2}$, respectively. Each state in the green subsphere has a corresponding
state in the red subsphere, so that the sum of their Bloch vectors is equal to
$\vecrts$ (chosen here along the $z$ axis). The gray sphere is the Bloch
sphere associated with the target state $\rho$, which can be seen at the
blue dot.  In both cases shown $h = 0.4$. In the left figure $\rts=0.3$ was
used, so the subspheres are contained within the Bloch sphere; while in the right
$\rts=0.5$. Separable states, represented by bold curves, are available only when $\rts\geq h$ as in the right figure. See the main text for details.
\label{fig:esferas}}
\end{figure} 
A geometric depiction of the set $\Omega_{\vecrts}$ provides
mathematical and physical intuition. This is given by the locus of all pairs of vectors $\vb{r}_1$ and $\vb{r}_2$ that lead to the same $\vecrts$ in \Eref{eq:bipartite CGM} as two spheres within the Bloch sphere. 
Notice that since the bipartite state is pure, purity of the two reduced states is the same, which implies that $ || \vb{r}_1 || = || \vb{r}_2 || = r$. Replacing $\vb{r}_1$ and $\vb{r}_2$ as in \Eref{eq:bipartite CGM} in the expression for purity equality, and completing the squares, we arrive to 
$ || \vb{r}_{1,2} -   \vb{c}_{1,2} || = R_{1,2} $ where $\vb{c}_1 = -\vecrts (1-h)/2h$ and $\vb{c}_2 = \vecrts (1+h)/2h$, are vectors indicating the position of the centers of the spheres, while $R_1 =
\rts(1+h)/2h$ and $R_2 = \rts(1-h)/2h$ correspond to their
radii, where $\rts \equiv ||\vecrts||$.  The two spheres touch in the target state, since $\vb{c}_{1,2} \pm R_{1,2}\vecrts/||\vecrts|| = \vecrts$. A geometrically more intuitive description of vectors $\mathbf{r}_i$ is described in \Aref{FSmeasure}.

Moreover, the radii of the spheres can vary from zero (for $h \to 1$) to infinity (for $h \to 0$); the spheres can be totally contained in the Bloch sphere, or be partially outside of it. Since a vector corresponding to a state must have norm less than one, this means that the sphere will be incomplete. This transition occurs when $\rts=h$. Separable states arise as solutions only when $\rts \geq h$. In \Fref{fig:esferas} we illustrate the above discussion.\par

To continue, we calculate the volume $V(\Omega_\epsilon) \equiv V_{\Omega_\epsilon}$ by integrating the FS measure over $\Omega_\epsilon$. In the limit where $V_\epsilon$ is an infinitesimal neighborhood centered at $\vecrts$, we find that
\begin{equation}
	V_{\Omega_\epsilon} = \frac{3 V_\epsilon}{2\pi(1+h)}
    \begin{cases}
    \frac{1}{h} & \text{if } 0 \leq \rts < h, \\[5pt]
   \frac{1}{1-h} \frac{1-\rts}{\rts} & \text{if } h \leq \rts \leq 1
    \end{cases}
    \, .
\label{eq:Volume-2qubits}
\end{equation}
Note that the preimage volume is constant when $0 \leq \rts < h$ (which corresponds to the two subspheres in ~\Fref{fig:esferas}  being entirely contained 
in the outer  Bloch sphere) and then decreases to zero, as $\rts$ tends to 1.
Thus, as $h$ approaches 1 (\ie, as $p$ tends to zero), the preimage volume
becomes constant, independent of the magnitude of $\rts$. 

\begin{figure}[b] 
\includegraphics[width=.9\linewidth]{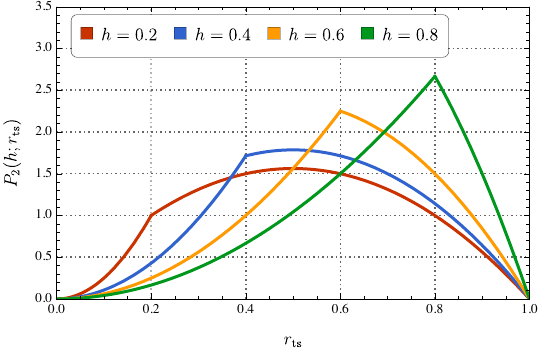}
\caption{
Probability density $P_2(h;\rts)$ \emph{vs.} $\rts$, for various values of
$h$. In all cases, $P_2$ tends to zero for both $\rts \rightarrow 0$
(maximally mixed state) and $\rts \rightarrow 1$ (pure states).
The cusp at $\rts=h$ corresponds to the two subspheres in ~\Fref{fig:esferas}
touching the outer Bloch sphere.
} 
\label{fig:P2Plot}
\end{figure}

In the case of separable states we find that 
\begin{equation}
	V_{\Omega_\epsilon^\otimes}=
\frac{V_\epsilon}{2\pi(1-h^2)}
    \begin{cases}
    0 
    & 
    \text{if } 
    0 \leq \rts < h
    \\[5pt]
   \frac{1}{\rts}
    & 
    \text{if } 
    h \leq \rts \leq 1
    \end{cases}
    \, .
\label{eq:Volume-2qubits-sep}
\end{equation}
Note that in the interval $0 \leq \rts < h$ the subspheres do not intersect the Bloch sphere and $\Omega_\epsilon^\otimes$ is the empty set.
The details of the above volume calculations can be found in \Aref{sec:Volumen}.
As expected, due to the covariance of the CG map established in \Eref{eq:simetria-map} and the unitary invariance of the ensemble of bipartite states, in both cases the preimage volume does not depend on the orientation of the target state.
Then, the relevant quantities are the probability densities $P_2$ and $P_2^\otimes$ of $\rts$, which are obtained from
the above results, respectively, by integrating the density $V_{\Omega_{\epsilon}}/V_{\epsilon}$ over the surface of constant $\rts$ in the Bloch sphere
\begin{align}
P_2(h;\rts)
&=
	 \frac{6 }{1+h}
    \begin{cases}
    \frac{1}{h} \rts^2 & \text{if } 0 \leq \rts < h, \\[5pt]
    \frac{1}{1-h} \rts (1-\rts)  & \text{if } h \leq \rts \leq 1
    \end{cases}
    \, ,
    \label{eq:PDF-2qubits}
    \\
P_2^\otimes(h; \rts)
&=
    \begin{cases}
    0 
    & 
    \text{if } 
    0 \leq \rts < h
    \\[5pt]
   \frac{2}{1-h^2} \, \rts
    & 
    \text{if } 
    h \leq \rts \leq 1
    \end{cases}
    \, .
\label{eq:PDF-2qubitsSep}
\end{align}
A plot of $P_2(h;\rts)$, for various values of $h$, appears in Fig.~\ref{fig:P2Plot}.

A couple of remarks are due at this point. The first one, is on
\Eref{eq:Volume-2qubits}, given that as $h$ goes to zero, it appears that
$V_{\Omega_\epsilon}$ increases boundlessly as $\rts <h$. This would be
problematic, as $V_{\Omega_\epsilon}$ is a subset of $\mathbb{P}$, and the
volume of the latter is normalized to 1. However, what really happens here is the
following: $V_{\Omega_\epsilon}$ is a well-defined function of the volume $V_\epsilon$ regardless its geometry (\eg, spherical, cubic, \etc). Since $V_{\Omega_\epsilon}(V_\epsilon=0)=0$, the Taylor expansion of
$V_{\Omega_\epsilon}(V_\epsilon)$, starts with the linear term,
$V_{\Omega_\epsilon}(V_\epsilon)=\partial V_{\Omega_\epsilon}/\partial
V_\epsilon|_{V_\epsilon=0} V_\epsilon + \mathcal{O}(V_\epsilon^2)$, and it is
exactly this truncated Taylor expansion what is provided
by \Eref{eq:Volume-2qubits}. When $h$ tends to zero, the above Taylor
expansion breaks down (just like, \eg, that of $\sqrt{x}$ does at $x=0$),
and \Eref{eq:Volume-2qubits} is no longer a good approximation to
$V_{\Omega_\epsilon}(V_\epsilon)$, which is always finite. In \Aref{sec:Volumen}, we explicitly compute the preimage volume of the origin $\rts = 0$ in the limit $h \rightarrow 0$.

The second remark is that \Eref{eq:Volume-2qubits-sep} has been derived
assuming that separable states are produced following the
distribution in \Eref{FSdmuprod}, which is only defined on the subset of product states.\par 

\subsubsection{Coarse-graining statistics via RMT methods} 
The methods developed in the previous sections allowed us to gain deeper physical insight into the problem. However, when considering systems composed of a larger number of particles, the number of parameters involved in the parametrization grows exponentially, which increases computational complexity. For this reason, we resorted to random matrix theory, which will subsequently enable us to analyze the case in which an $N$-partite system is reduced to a single one.
To introduce the RMT framework, in this subsection we compute the probability density of $\rho$ for the bipartite case and compare our results with those obtained using geometric methods.

To establish the connection between the statistical properties of $\rho$ and those of its matrix elements, we first recall the derivative principle for unitarily invariant ensembles of matrices of dimension $D \times D$. This theorem relates the joint probability density function (PDF) $f_{\rho}$ of its eigenvalues $\lambda_i$ and the joint PDF $\Psi_{\rho}$ of its diagonal elements. Subsequently, we will employ this relationship in the case where the target state is obtained from pure bipartite states through the coarse-graining mapping. Since the eigenvalues of a single-qubit state depend only on $\rts$, the PDF of the eigenvalues directly gives us the PDF of $\rts$. We will express the joint PDF of the diagonal elements of the target state in integral form, considering the conditions that these elements must satisfy. 
The resulting integrals are then evaluated in {\it Laplace space}, which naturally facilitates the generalization of the expressions to $N$-partite systems.  Finally, the Laplace transforms are inverted, and $\Psi_{\rho}$ is substituted into the derivative principle theorem to obtain $f_{\rho}$. 

The derivative principle was developed in \cite{Christandl2014}, where it was used to calculate the 
eigenvalue PDFs of reduced density matrices of multipartite entangled states. Using the standard language of random matrix theory, it is enounced as follows \cite{Mejia2017,Kieburg2023}:

\textit{Let $\rho$ be a random matrix drawn from a unitarily invariant random matrix ensemble, $f_{\rho}$ the joint PDF of the eigenvalues of $\rho$, and $\Psi_{\rho}$ the joint PDF of the diagonal elements of $\rho$. Then,}
\begin{equation}
    f_{\rho}(\vec{\lambda}) = 
    \left(\prod_{n=1}^{D}n!\right)^{-1} \Delta(\vec{\lambda})\Delta(-\partial_{\vec{\lambda}})\Psi_{\rho}(\vec{\lambda}), 
    \label{eq:appF-Derivative-principle}
\end{equation}
\textit{where $\vec{\lambda} = (\lambda_{1}, \lambda_{2}, \dots, \lambda_{D})$ are the eigenvalues of $\rho$, $\Delta(\vec{\lambda}) = \prod_{i<j} (\lambda_{j} - \lambda_{i})$ is the Vandermonde determinant, and $\Delta(-\partial_{\vec{\lambda}})$ is the differential operator $\prod_{i<j}\big(\frac{\partial}{\partial\lambda_{i}} - \frac{\partial}{\partial\lambda_{j}}\big)$.
}

In this article, we have restricted ourselves to target states that are monopartite, so in all cases we have $D=2$. We identify the diagonal elements of $\rho$ as $\rho_{00}$ and $\rho_{11}$. Since $\rho$ has unit trace, if the PDF of $\rho_{00}$ is known, the joint PDF of the diagonal elements of the target state can be expressed as $\Psi_{\rho}(\rho_{00},\rho_{11})=\Psi(\rho_{00})\delta(\rho_{00}+\rho_{11}-1)$. Substituting $\Psi_{\rho}(\rho_{00},\rho_{11})$ into \Eref{eq:appF-Derivative-principle}, it follows that $f_\rho(\lambda_1,\lambda_2) = \frac{1}{2}(\lambda_2-\lambda_1)\Psi'(\lambda_1)\delta(\lambda_1+\lambda_2-1)$. Integrating the PDF of the eigenvalues over $\lambda_2$, we obtain the distribution of the single independent eigenvalue $f_\rho(\lambda_1) = \frac{1}{2}(1-2\lambda_1)\Psi'(\lambda_1)$. Moreover, the eigenvalues of the target state $\rho$ are $\tfrac{1}{2}(1\pm\rts)$, so there are two values of $\lambda_1$ that correspond to the same $\rts$. Since both branches contribute equally to the probability, performing the change of variable $\lambda_1 = \tfrac{1}{2}(1+\rts)$ we finally obtain
\begin{equation}
	f_\rho(\rts) = -\rts \frac{\dd\Psi(\rts)}{\dd{\rts }}\, .
	\label{eq:PDF r}
\end{equation}
We will now use the relation (\ref{eq:PDF r}) in the simplest case, when $N=2$. Given the structure of the coarse-grained map $\mathcal{C}$, when applied to the pure bipartite state $\ket{\psi}=\sum_{i,j}c_{ij}\ket{i,j}$, the diagonal elements of the target state have the form:
\begin{align} 
	\rho_{00} &= x_{00} + p_{1}x_{01} + p_{2}x_{10},
	\label{appF:diagonal-elements-target-0} \\  
	\rho_{11} &= x_{11} + p_{2}x_{01} + p_{1}x_{10} = 1-\rho_{00},
	\label{appF:diagonal-elements-target-1}
\end{align}  
where it holds that $p_{1}+p_{2}=1$ and $x_{ij}=\abs{c_{ij}}^{2} \in [0,1]$. Additionally, since the state $\ket{\psi}$ is normalized, the condition $\sum_{i,j}x_{ij}=1$ is satisfied, which implies that $\rho_{00}+\rho_{11}=1$.

The equations (\ref{appF:diagonal-elements-target-0}) and (\ref{appF:diagonal-elements-target-1}) define a mapping from the $3$-simplex in the space of the $x_{ij}$'s to the $1$-simplex in the space of the $\rho_{ii}$'s. It can be shown that, for density matrices distributed according to the FS measure, the variables $x_{ij}$ follow a uniform distribution on the corresponding $3$-simplex $S_0 \subset \mathbb{R}^4$ \cite{Zyczkowski1999}. Therefore, the probability measure on $S_0$ can be expressed as $\dd\nu = 3! \dd x_{00}\dd x_{01}\dd x_{10}$, using coordinates $(x_{00},x_{01},x_{10})$ with $0 \leq x_{ij}$ and the constraint $0\leq 1-x_{00}-x_{01}-x_{10}$.

The PDF induced in the space of the $\rho_{ii}$'s is obtained by integrating the induced measure over the preimage of a point in the $1$-simplex. To perform this integration, we take $\rho_{00}$ as the coordinate in both the $1$-simplex and the $3$-simplex, since $\rho_{00}$ is defined in terms of the coordinates $x_{ij}$, as shown in  \Eref{appF:diagonal-elements-target-0}. In this way, the preimage of $\rho_{00}$ is parametrized by $x_{01}$ and $x_{10}$, then
\begin{align}
	\Psi(\rho_{00}) &= 3!\int_H \dd x_{01}\dd x_{10} \notag \\  
	&= 3!\int_{\mathbb{R}^{2}_{+}} \Theta(\rho_{00}-\tilde{\rho}_{00})\Theta(\rho_{11}-\tilde{\rho}_{11})\dd x_{01}\dd x_{10} \notag \\ 
	&\equiv 3! A(\rho_{00},\rho_{11})
\end{align}  
where $H$ represents the preimage of $\rho_{00}$, defined as the region where the inequalities $0\leq x_{01}$, $0\leq x_{10}$, $0\leq \rho_{00}-\tilde{\rho}_{00} = x_{00}$, and $0\leq \rho_{11}-\tilde{\rho}_{11} = x_{11}$ hold, with  
\begin{align} 
	\tilde{\rho}_{00} &= p_{1}x_{01} + p_{2}x_{10}, \label{appF:diagonal-elements-target-0-tilde} \\  
	\tilde{\rho}_{11} &= p_{2}x_{01} + p_{1}x_{10}. \label{appF:diagonal-elements-target-1-tilde}
\end{align}

There are several ways to calculate the area $A(\rho_{00},\rho_{11})$ bounded by the lines $\rho_{ii} = \tilde{\rho}_{ii}$ in the $x_{01}$-$x_{10}$ space. However, with the aim of simplifying the calculations and generalizing the procedure to $N$-partite systems, we propose reformulating the problem in the Laplace space. For this, we consider the variables $\rho_{ii}$ as independent in the domain $\rho_{ii}\in [0,\infty)$, then we can calculate the Laplace transform of $A(\rho_{00},\rho_{11})$ and finally perform the integrals in $x_{01}$ and $x_{10}$. As a result, we obtain
\begin{align}
	\tilde{A}(s_{0},s_{1}) &= \mathcal{L}_{s_{0},s_{1}}\left[A(\rho_{00},\rho_{11})\right] \nonumber \\  
	&= \frac{1}{s_{0}s_{1}(s_{0}p_{1}+s_{1}p_{2})(s_{0}p_{2}+s_{1}p_{1})}.
	\label{appF:Space-Laplace}
\end{align}
\begin{figure}[t]
\center
			\includegraphics[width=.7\columnwidth]{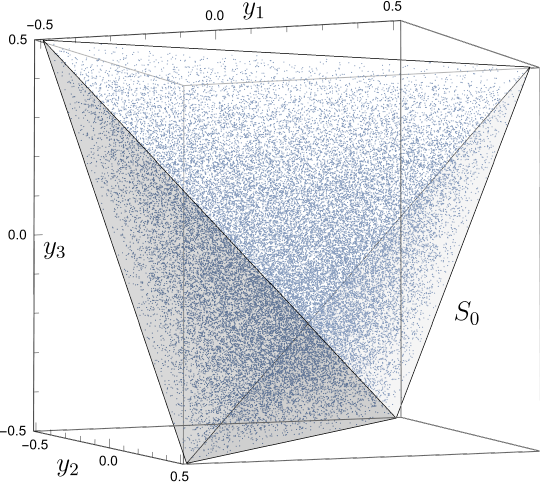}
			\caption{The simplex $S_0$ and the image of 40,000 bipartite states in it.}
			\label{fig:simplexPPlot}
\end{figure}		
Next, we calculate the inverse Laplace transform of the previous equation to obtain the PDF of $\rho_{00}$:
\begin{align}
\Psi(\rho_{00}) &= 3! A(\rho_{00},\rho_{11}=1-\rho_{00})
\nonumber
\\
&=
\frac{3}{(p_2-p_1)p_1 p_2}
\bigg( (p_2-p_1)(1-\rho_{00})^2
\nonumber
\\
& 
+(p_1-\rho_{00})^2 \Theta\qty(1-\frac{\rho_{00}}{p_1})
\nonumber
\\
& \left.
-(p_2-\rho_{00})^2 \Theta\qty(1-\frac{\rho_{00}}{p_2}) 
\right)
\, .
\label{Psia}
\end{align}

Evaluating the function in (\ref{Psia}) at $\frac{1}{2}(1+\rts)$, and substituting into \Eref{eq:PDF r} with $p_1=\frac{1}{2}(1-h)$, $p_2=\frac{1}{2}(1+h)$ and $h\in[0,1]$, we obtain
\begin{equation}
f_\rho(\rts) = \frac{6 \rts}{h(1-h^2)}\qty[h(1-\rts)+(\rts-h)\Theta(h-\rts)] \, ,
\end{equation}
which recovers the PDF of $\rts$ obtained in \Eref{eq:PDF-2qubits} from the previous section.
\subsubsection{A geometric interlude\label{Agi}} 
The techniques used in the previous section  are both powerful and intuitively opaque (at least to the authors). We try here to  gain an intuitive, geometric understanding of some of the previous results. 

Consider $\mathbb{R}^4$ with coordinates $\{x_{00},x_{01},x_{10},x_{11}\}$, then the fact the $x_{ij}$'s are non-negative and sum to one means that the locus of $x$ is the simplex $S_0$ with vertices on the standard basis vectors, and $a \equiv \rho_{00}=r'_1 \cdot x$, with $r'_1=(1,p_1,p_2,0)$, see~\Eref{appF:diagonal-elements-target-0}. It will prove convenient to switch to coordinates $(y_1,y_2,y_3,y_4)$, defined by $y=M x$, with $M \in SO(4)$ given by
\begin{equation}
M=\frac{1}{2}
\left(
\begin{array}{cccc}
1 & -1 & 1 & -1
\\
1 & -1 & -1 & 1
\\
1 & 1 & -1 & -1
\\
1 & 1 & 1 & 1
\end{array}
\right)
\, .
\end{equation}
Then $a=r_1 \cdot y$, with $r_1=M r'_1=(p_2,0,p_1,1)$ --- note that we denote vectors in the $x$-basis by a prime, and those in the $y$-basis without. The hyperplane $\Pi_0$
of $S_0$ has the equation $x \cdot n'=1/2$, with $n'=(1,1,1,1)/2$ being the unit
normal vector of $\Pi_0$. In the $y$-coordinates that equation becomes $y \cdot
n=1/2$, with $n=M n'=(0,0,0,1)$, \emph{i.e.}, all points on $S_0$ have
$y_4=1/2$, so we may truncate $y$ to its first three components when dealing
with $S_0$, $y=(\vec{y},1/2)$ on $\Pi_0$. The vertices of $S_0$, in the
$\vec{y}$-coordinates,  are given by the first three entries of the columns of
$M$,
\begin{multline}
\label{S0verty}
S_0=
\bigg\{
 \frac{1}{2}(1,1,1),\frac{1}{2}(-1,-1,1),\frac{1}{2}(1,-1,-1), \\
 \frac{1}{2}(-1,1,-1) 
 \bigg\}
\, .
\end{multline}

The preimage of $a$ on $S_0$ consists of those points $\vec{y}$ satisfying
\begin{equation}
\label{preimagea}
\vec{r}_1 \cdot \vec{y}=a-\frac{1}{2}
\, ,
\end{equation}
with $\vec{r}_1=(p_2,0,p_1)$. The above equation defines a 2-plane $\Pi_{p_1 a}$ in $\vec{y}$-space, which, generically,  intersects $S_0$ in a convex polygon $H_{p_1 a}$. The value of $p_1$ determines the orientation of $H_{p_1 a}$, while varying $a$ shifts $H_{p_1a}$ parallel to itself. As mentioned before, when $\rho_{\Psi}$ follows the FS distribution, the corresponding distribution of $\vec{y}$ is uniform on $S_0$. In Fig.~\ref{fig:simplexPPlot} we plot $S_0$ and the image of 40,000 bipartite states in it, generated with the FS measure. 

\begin{figure}[htbp]
\center
			\includegraphics[width=.8\columnwidth]{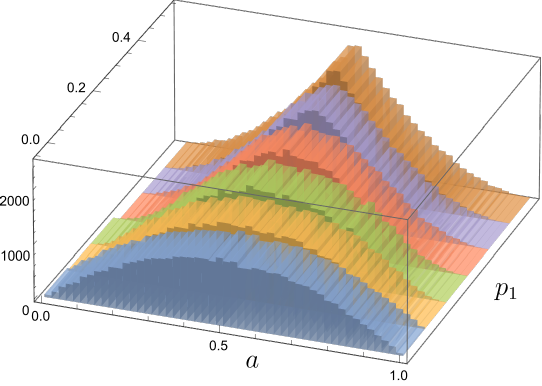}
			\caption{Histogram of the distribution of $a \equiv \rho_{00}$, for various values of $p_1$. Note that for $p_1=.5$ (orange/ochre histogram in the back) there is a cusp at $a=.5$, which is smoothed out for the other values of $p_1$.}
			\label{fig:H3DPlot}
\end{figure}
\begin{figure}[b]
\center
			\includegraphics[width=\columnwidth]{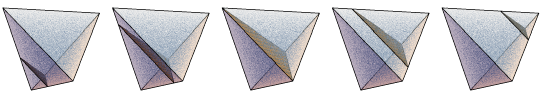}\\
				\includegraphics[width=\columnwidth]{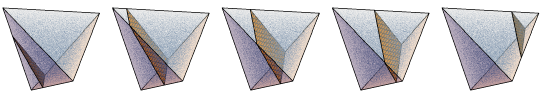}
			\caption{\textbf{Top row:} The preimage $H_{p_1a}$ of $a$ in $S_0$, for $p_1=.5$ and $a=.2$, $.4$, $.5$,  $.6$, $.8$ (from left to right). Compare with the orange/ochre histogram in Fig.~\ref{fig:H3DPlot}.  Because of the particular orientation of $H_{p_1 a}$ in this case, its size increases linearly with $a$, which implies that the ``tails'' on either side of the cusp in Fig.~\ref{fig:H3DPlot} are parabolic, while the cusp itself corresponds to the frame in the middle. \textbf{Bottom row:} Same as above, for $p_1=.3$. Because the orientation of $H_{p_1 a}$ in this case, \emph{w.r.t.} $S_0$, is generic, the cusp observed for $p_1=.5$  is smoothed out --- nonanalyticity still creeps in when, \emph{e.g.}, $H_{p_1a}$ meets (or leaves) the ``front'' edge of $S_0$.}
			\label{fig:gridsimplexPlot05}
\end{figure}

\begin{figure}[t]
\center
			\includegraphics[width=.8\columnwidth]{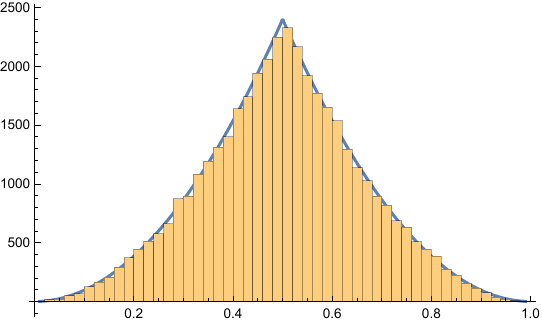}
			\caption{PDF  of $a$ (appropriately scaled), for $p_1=.499$, superimposed on the corresponding histogram  (orange/ochre histogram in the ``back'', in Fig.~\ref{fig:H3DPlot}).}
			\label{fig:aPDFhistPlot}
\end{figure}

		
Then the PDF of $a$, for given $p_1$, is proportional to the area of $H_{p_1 a}$, since, due to linearity, the ``thickness'' of the preimage of an infinitesimal segment $\text{d}a$ is constant, independent of $a$. In Fig.~\ref{fig:H3DPlot} we plot histograms of the distribution of $a$ for various values of $p_1$, using the above mentioned set of states.
In Fig.~\ref{fig:gridsimplexPlot05} we plot $H_{p_1 a}$ for two values of $p_1$, and, for each of them, for  six values of $a$. As explained in the caption to the figure, the evolution of the area of $H_{p_1 a}$ as $a$ varies, reproduces  the characteristics of the histograms in Fig.~\ref{fig:H3DPlot}.
To check on our results of the previous section, we compare with the expression (\ref{Psia}) verifying the (piecewise) quadratic dependence on $a$, as expected from the geometric considerations presented above. A (suitably scaled) plot of $\Psi(\rho_{00})$, is superimposed on the corresponding histogram in Fig.~\ref{fig:aPDFhistPlot}.
\subsection{$N$-partite systems} 
\label{sec:N:partite:volumen}

A geometric description of systems with $N>2$ qubits becomes increasingly
complex because the number of parameters needed for a complete parametrization
grows like $2(2^{N}-1)$. Instead, we resort to the derivative principle theorem
of RMT \citep{Christandl2014} which establishes a relationship, shown in \Eref{eq:appF-Derivative-principle}, between the joint PDF $\Psi_{\rho}$ of the diagonal elements of a state $\rho$ and $f_\rho$ the joint PDF of its eigenvalues, provided that $\rho$ belongs to a unitarily invariant ensemble of matrices of dimension $D\times D$. 

Let us express pure $N$-partite states as $\ket{\psi}=\sum_{l} c_{l}
\ket{l}$ with $\ket{l}$  an element of the
computational basis. We can then calculate $\Psi_{\rho}(\rts)$, and using \Eref{eq:appF-Derivative-principle},  we find that
the PDF of finding a target state, defined by the Bloch vector $\mathbf{r}_{\rmt}$, when applying the CG map to an $N$-partite state, is
\begin{multline}
P_N(p,\rts) =\\
c_N \rts
\sum_{l\neq (0,\dots,0)}\frac{ \left( \Sigma_i \tilde p_i^{(l)} - \rts \right)^{2^ {N}-3}
\Theta\!\left( \Sigma_i \tilde p_i^{(l)} - \rts\right) }
{\prod\limits_{l'\neq (0,\dots,0)} l'\cdot \tilde p^{(l)} }. 
\label{eq:Prob-Vol-Gen-N}
\end{multline}
where $c_N=(2^{N}-1)(2^{N}-2)/2^{2^{N}-2}$ is a normalization constant, $l=(l_1,\ldots,l_{N})$ is a multiindex with $l_i=0,1$, $p=(p_1,\ldots,p_{N})$ is the probability vector of the CG map, and $\tilde p_i^{(l)} = (2l_i -1) p_i$ are the components of the probability vector with signs flipped according to $l$.
The explicit calculation can be found in \Aref{RMT-Method-N}.
\Eref{eq:Prob-Vol-Gen-N} is piecewise continuous and, as a function of $p$, 
has poles at those points where two or more probabilities $p_i$ are equal to each other.
Furthermore, \Eref{eq:Prob-Vol-Gen-N} reduces to \Eref{eq:Volume-2qubits} when $N=2$,
which supports its validity.

\begin{center}
\begin{figure}[htbp]
\includegraphics[width=1\linewidth]{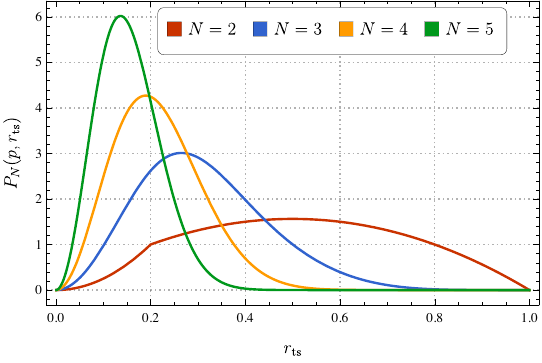}
\caption{PDF for N=2, 3, 4, and 5. Each of the functions has been normalized to the Bloch sphere of the target state.
The probabilities $p$ used are: For $N=2$, $\{0.4,0.6\}$; for $N=3$, $\{0.4, 0.35, 0.25\}$; for $N=4$, $\{0.4, 0.35, 0.15, 0.1\}$; and for $N=5$, $\{0.4, 0.35, 0.15, 0.08, 0.02\}$. Note that as the number $N$ of qubits in the system increases, the volume tends to take the form of a Gaussian. \label{fig:PDF-N}}
\end{figure}
\end{center}

\Fref{fig:PDF-N} displays the corresponding PDF for the radius $\rts$ for $2$, $3$,
$4$, and $5$ qubits for some specific values of the probabilities $p_i$.
Note that as the number of qubits in the system increases, 
the probability density progressively resembles a Gaussian distribution, with mean value and variance that tend to zero, so that, for $N$ large enough, the observed state is the maximally mixed one. 
\section{Average state \label{sec:average-state}} 
Now we proceed to calculate the average states of the sets
$\Omega_\epsilon$ and $\Omega_{\epsilon}^{\otimes}$, describe their symmetries, and discuss some special cases.
Let us recall that, within our framework, these average states serve as the preimages of the target state in the sense that $\overline{\mathcal{C}^{-1}}(\rho)=\varrho^{\avs}$. 
This average state plays a central role in mediating the connection between the emergent coarse-grained dynamics 
and the fine-grained unitary dynamics. 
Furthermore, we examine how the purity and coherence of the average states depend on the target radius $\rts$. 
Finally, we comment on the difficulties involved in computing average states for $N$-partite systems and outline the strategy we adopt to address them.

The average state $\varrho^{\avs}$ of the set $\Omega_{\epsilon}^{N}$ is defined as
\begin{equation}
	\varrho^{\avs} = \frac{1}{V(\Omega_{\epsilon}^{N})}\int_{\Omega_{\epsilon}^{N}} \dd\mu\, \varrho
    \, ,
	\label{eq:def AS1}
\end{equation} 
and may be expanded in tensor products of Pauli $\sigma_\mu$ matrices, $\mu=0,1,2,3$, (with $\sigma_{0}=I$),
\begin{equation}
    \label{varrhoexpands}
    \varrho^{\avs}=\frac{1}{2^N}\sum_{\nu}\varrho^{\avs}_{\nu}\sigma_{\nu}
    \, ,
\end{equation}
where $\nu=(\nu_{1}\ldots \nu_{N})$ is a multiindex, and $\sigma_\nu \equiv \sigma_{\nu_1} \otimes \ldots \otimes \sigma_{\nu_N}$. Since the Pauli matrices satisfy $\tr(\sigma_{\mu_1} \sigma_{\mu_2}) = 2\delta_{{\mu_1}{\mu_2}}$, the coefficients $\varrho_\nu$ can be obtained as
\begin{equation}
	\varrho^{\avs}_\nu = \tr(\varrho^{\avs}\sigma_\nu) = \frac{1}{V(\Omega_{\epsilon}^{N})}\int_{\Omega_{\epsilon}^{N}} \dd\mu \ev{\sigma_\nu}_{\varrho}\, .
    \label{eq:varrho_nu}
\end{equation}

For bipartite states ($N=2$), this calculation involves the 16 elements of the
density matrix, but only 6 elements are non-zero, with
$\varrho^{\avs}_{0,0}=1$. Additionally, we find that $\varrho^{\avs}_{2,2}=\varrho^{\avs}_{1,1}$, so
only 4 elements are needed to fully determine the average state. This
simplification arises from the symmetry of the system, as $\rho$ is
polarized in the z-direction. Detailed calculations of the average state
components for the set $\Omega_\epsilon$ and a visualization of these
components are provided in \Aref{Average-state}.

For the average state of the set $\Omega_\epsilon$, when $\rts\leq
h$, we find that also $\varrho^{\avs}_{3,3}=\varrho^{\avs}_{1,1}$, meaning that in
reality, knowing $3$ average expected values is sufficient to completely
determine $\varrho^{\avs}$. When $\rts>h$, the $4$ average expected values are
distinct from each other, but it can be identified that there exists a symmetry
$\varrho^{\avs}_{0,3}(h,\rts)=\varrho^{\avs}_{3,0}(-h,\rts)$, which is also valid for the
average state of the set of product bipartite states. 
\par
In order to gain physical insight about the properties of the average state, we consider the two extreme cases: the maximally mixed and pure target states. 
If the target state is $\rho=\frac{1}{2}I$ (equivalently $\rts=0$), the state satisfies $U\rho U^{\dagger}=\rho$ for any unitary $U$. Then, the symmetry
property \Eref{eq:simetria-map} of the CG map ensures that the average state will satisfy the
condition $(U\otimes U) \varrho^{\avs} (U^{\dagger} \otimes U^{\dagger}) = \varrho^{\avs}$.
This implies that the average state will be isotropic and thus is of the form
\bq 
	\varrho^{\avs} = \alpha \ket{\phi_{1}}\bra{\phi_{1}} + \textstyle{\frac{1-\alpha}{4}}I\otimes I
    \, ,
	\label{eq:werner-form}
\eq 
where $\ket{\phi_{1}}$ is the singlet state~\cite{Werner1989}.
It turns out (using \Eref{eq:varrho} and \Eref{eq:ci aprox}, see \Aref{Average-state}) that $\alpha=\frac{1-h}{3(1+h)}$. Note that the preimage of the single-qubit  maximally mixed state is, by definition, the set of maximally entangled two-qubit states. 
On the other hand, the pure target state $\ket{\hat{\mathbf{n} }}$ has as its preimage (for generic $h$) the coherent state $\ket{\hat{\mathbf{n} }} \otimes \ket{\hat{\mathbf{n} }}$.

\par
We now turn our attention to the set $\Omega_{\epsilon}^{\otimes}$. This case is particularly interesting because the convexity of the set of separable mixed states ensures that the average of $\Omega_{\epsilon}^{N\otimes}$ is separable, meaning that it does not exhibit entanglement between the qubits that compose it. In this case, we find that the components of the average state satisfy \(\varrho^{\avs}_{3,3} = \varrho^{\avs}_{3,0}\varrho^{\avs}_{0,3}\), showing that the correlation along the z direction factorizes into the local polarizations of each qubit.
This average state only exists when $r_\rmt \geq h$, as in the opposite case, the set $\Omega_{\epsilon}^{\otimes}$ is empty. 
Notably, when $r_\rmt = h$, the first element of the set appears, specifically $\varrho^{\avs} = \ket{1,0}\bra{1,0}$. In this scenario, the average state associated with a target state is the same for generic $h$. Additionally, when $r_\rmt = 1$, the preimage of the
target state $\rho$ is the same for both the complete and separable
cases, i.e., $\Omega_{\vb{R}}^{\otimes} = \Omega_{\vb{R}}$.
\par
\begin{figure}[t]
\includegraphics[width=1\linewidth]{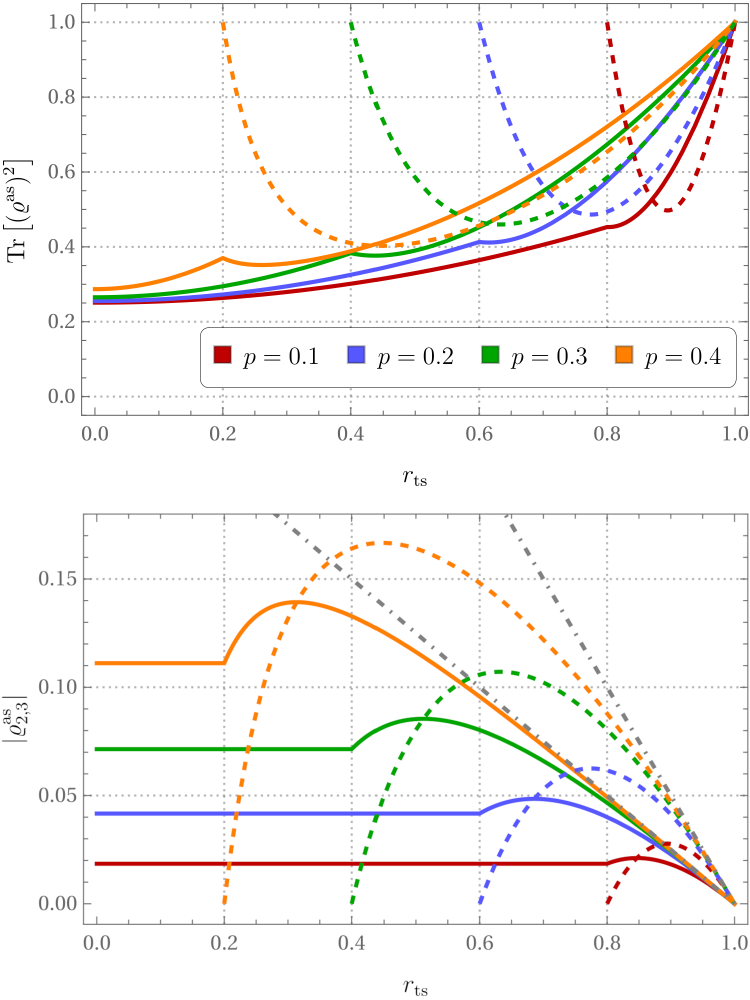}
\caption{
The top plot shows the behavior of purity of the average state $\varrho^{\avs}$
as a function of $\rts$ for different values of $p=1/2(1-h)$. 
The bottom plot shows how $\varrho^{\avs}_{2,3}=\varrho^{\avs}_{3,2}$ (the only coherences
in the computational basis that are not null) change with  $\rts$.
Both plots display results for $\Omega_{\epsilon}$, represented by solid lines, and  $\Omega^{\otimes}_{\epsilon}$, shown with dashed lines. The dotted gray lines correspond, from left to right, to $\textstyle{\frac{1}{4}}(1-\rts)$ and $\onehalf(1-\rts)$.
\label{fig:propiedades}
}
\end{figure}

In \Fref{fig:propiedades}, the purity of the average state as a
function of $\rts$ is depicted, along with the behavior of the coherences.
Notice that the density matrix of the average state contains only two non-zero coherences. 
In the computational basis, these are $\varrho^{\avs}_{01,10} \equiv \varrho^{\avs}_{2,3}$ 
and $\varrho^{\avs}_{10,01} \equiv \varrho^{\avs}_{3,2}$ ---not to be confused 
with the $(2,3)$ and $(3,2)$ elements in the Pauli basis, which are zero---, and they are equal 
$\varrho^{\avs}_{2,3} = \varrho^{\avs}_{3,2}$.
We observe that the average state derived from the set
$\Omega_{\epsilon}$ is pure only when the target state is pure; however,
the probability density of measuring this state is zero. Furthermore, the coherences of
this average state remain constant in the region where $\rts \leq h$, and then
they tend towards the line $\frac{1}{4}(1-\rts)$; in fact, they are
bounded: $0\leq \abs{\varrho^{\avs}_{2,3}}\leq \frac{1}{4}(1-\rts)$. Regarding the
average state
obtained from the set $\Omega_{\epsilon}^{\otimes}$, there are two values of
$\rts$ for which the average state is pure, both with a finite probability density of
being observed. When $\rts= h$ the average state is $\ket{\psi}=\ket{1,0}$
and when $\rts=1$, it becomes $\ket{\psi}=\ket{0,0}$. In both cases,
the coherences vanish. Accordingly, the off-diagonal term satisfies the bound $0\leq \varrho^{\avs}_{2,3} \leq \onehalf(1-\rts)$. 
\par
Calculating the average state for $N>2$ with geometrical arguments
is again difficult, for identical reasons as exposed in
\Sref{sec:N:partite:volumen}. 
Additionally, since the sets of average expected values $\varrho^{\avs}_{i,j}$ are not unitarily invariant ensembles, the RMT methods used for the volume of $N$-partite systems are not applicable for calculating the corresponding expected values.
Nevertheless, an analysis of the average state of the set $\Omega_{\epsilon}$ has revealed 
that the symmetry $\mathcal{C}[(U\otimes U)\ket{\psi}\bra{\psi}(U^{\dagger}\otimes U^{\dagger})]
=U\mathcal{C}[\ket{\psi}\bra{\psi}]U^{\dagger}$ 
provides insight into the general structure of that state. Consequently, the
analysis of the symmetry in the \Eref{eq:simetria-map} represents a promising approach
for understanding the average state of a set composed of $N$-partite states.
\section{Statistics}\label{sec:statistics} 
\subsection{Numerical preimage volumes} 
\label{sec:num_preimage}

As discussed before, under a CG map bipartite systems of two qubits are mapped to
single qubit mixed states $\rho \in \mathcal{M}_2$. In this particular
case, the manifold $\mathcal{M}_2$ is diffeomorphic to the Euclidean ball
$B^3$, which allows us to use the standard Euclidean metric to specify a
neighborhood $V_\epsilon$ around the target state $\rho$ parametrized by the
vector $\vecrts$. We calculate the volume $V(\Omega_\epsilon)$ of the preimage --- lying in the space of pure bipartite states $\mathbb{C}P^3$--- of $\vecrts$.

If we want to address the more general situation in which the CG map sends
$N$-partite pure states to $m$-partite mixed states $\rho \in
\mathcal{M}_{D}$, where $\mathcal{M}_{D}$ is the space of Hermitian positive
matrices of size $D = 2^m$, is given
\bq
	\mathcal{C}_m[\varrho] = \sum_{i=1}^b p_i \tr_{c(i)}\varrho\,,
\eq
where $b=\binom{N}{m'}$, $c(i)$ is the $i$-th element of the set of
$m'$-combinations from $N$ elements and $m+m'=N$. We notice that the target
space is no longer diffeomorphic to any subspace of the Euclidean space,
furthermore, the dimensions of both spaces $\mathbb{C}P^{2^N-1}$, and
$\mathcal{C}_m(\mathbb{C}P^{2^N-1})\subset \mathcal{M}_D$ where the target
state lives, grow large at a very quick rate.

An alternative for exploring the behavior of the CG map in such cases is to calculate the normalized expected value of the composite spin operator, which correspond to the natural action of the elements of the Lie algebra $\mathfrak{su}(2)$ on the tensor product of spin states 
\bq
		\mathcal{S}_{\alpha} = \sum_{i=1}^m I_1\otimes\cdots\otimes \sigma_{\alpha_i}\otimes\cdots\otimes I_m\,,
	\label{eq:composite_spin_operator}
\eq
where the subindex in the operatores label the Hilbert space where the $i$-th qubit lives and $\sigma_{\alpha}$, $\alpha=1,2,3$ are the 3 Pauli matrices. Then, given a target state $\rho \in \mathcal{M}_{D}$, its normalized spin expectation value (SEV) can be computed as
\begin{equation}
	\langle \bmmcS{} \rangle = 
	     \frac{1}{m}\tr(\rho\bmmcS{} ) \equiv \vecrts\in B^3\,,
	\label{eq:normalized_SEV}
\end{equation}
which defines a map
$
\langle\bmmcS{}\rangle:
\mathcal{C}_m(
\mathbb{C}
P^{2^N-1})\rightarrow B^3
$, 
that composed with the CG map gives us a mapping that sends a pure $N$-partite state to a physical vector in $\mathbb{R}^3$.

Due to the spherical symmetry of the SEV probability distribution, we're only interested in the distribution of its modulus $\rts$. 
Then we will ask for the probability of an $N$-partite state to be
mapped to a spherical shell of radius $\rts$ and thickness $\epsilon$. This
question is equivalent to the one made before in the bipartite case, when we
asked for the probability of a two qubit system to be mapped to a sphere
$V_{\epsilon}\subset B^3$ of radius $\epsilon$, since the volume
$V\qty(\Omega_{\epsilon}^{N,m})$ of the preimage of some neighborhood around
$\vecrts$ does not depend on the geometry of $V_\epsilon$, but only on its
volume.\par

As it has been mentioned before in this work, to provide an analytic answer in the general case is complicated due to the large amount of parameters needed to specify both the original pure states of $N$ qubits and their images under the CG map. However, such analytic result is provided in \Sref{sec:N:partite:volumen} for the cases $N\rightarrow 1$, and here we will use Monte Carlo simulations with samples of $n=10^4$ random $N$-partite systems with $N=2,3,4,5$ to compare such results, and confirm that this statistical analysis can be used in the pursue of understanding the more general $N\rightarrow m$ case.

Then, we may call $W$ the set of all the SEV calculated as in \Eref{eq:normalized_SEV} from the sample of random target states. The preimage volume $V(\Omega^N_\epsilon)$ can be calculated through a Monte Carlo simulation as
\bq
V\qty(\Omega_\epsilon^{N})_{\text{MC}} = \frac{1}{n} \sum_{r\in W}\chi_{V_{\epsilon}}(r)\,,
\label{eq:numeric_preimage_vol}
\eq

where $\chi_{V_{\epsilon}}(r)$ is the characteristic function of $V_{\epsilon}$ that in this case is taken as a spherical shell of thickness $\epsilon$ centered in $\rts$,
\bq
   \chi_{V_{\epsilon}}(r) = 
   \bc 
   1 & |r - \rts| \leq \epsilon/2\\
   0 & \mathrm{else}
   \ec\,,
\eq 
the superindex in $\Omega_{\epsilon}^{N}$ denotes the number of qubits that constitute the original pure state, and the superindex $m=1$ has been dismissed for simplicity in the notation.
Then, we calculate the PDF $P_{N}(p,\rts)$ as
\bq
P_{N}(p,\rts) = \frac{V(\Omega_\epsilon^N)}{V_\epsilon}4\pi\rts^2\,,
\eq
where the factor $4\pi \rts^2$ is due that $V(\Omega_\epsilon^N)/V_\epsilon$ represents the PDF of a target state specified only by the module of its SEV, then, since this probability does not depend on the orientation of the target state, we can obtain the PDF of the complete set of all states with the same $\rts$, which is the quantity given by $P_N(p,\rts)$, just by multiplying by the area of the sphere with such radius.

\begin{figure}[htbp]
\centering
  \includegraphics[width=1\linewidth]{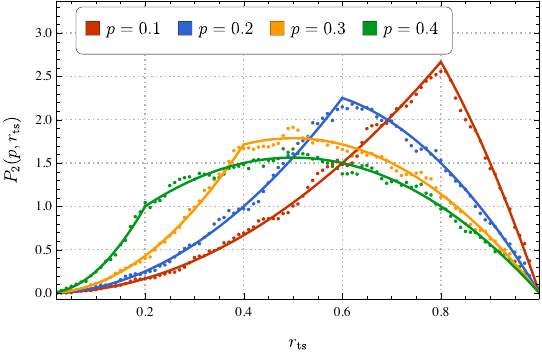}
  \caption{Plot of the PDF function $P_2(p,\rts)$ taking different values of $p = 1/2(1-h)$ (solid lines), compared with the results obtained numerically using a Monte Carlo simulation (dots), with $\epsilon=0.04$.}
  \label{fig:2to1_compare_PDF}
\end{figure}

\begin{figure}[htbp]
\centering
  \includegraphics[width=\columnwidth]{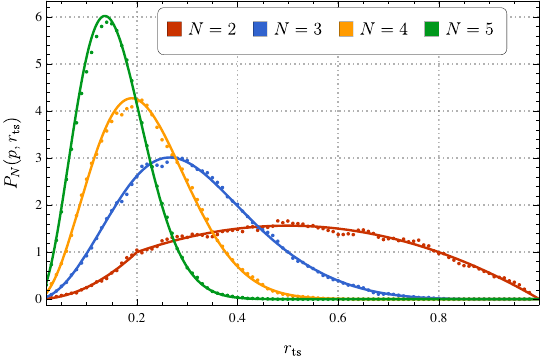}
    \caption{Plot of the PDF, $P_N(p,\rts)$ for the cases when $N=2$ to 5 (solid lines), compared with the results obtained numerically using a Monte Carlo simulation (dots), with $\epsilon = 0.04$.}
  \label{fig:PDF_compare}
\end{figure}

As expected, when we performed the numerical experiment in the case when $N=2$, we recovered the same results as with \Eref{eq:Volume-2qubits} and the corresponding case of \Eref{eq:Prob-Vol-Gen-N}, shown in \Fref{fig:2to1_compare_PDF}. Similarly, for the cases when $N=2,3,4,5$, as shown in \Fref{fig:PDF_compare} the results are consistent with the analytic expression obtained in \Eref{eq:Prob-Vol-Gen-N} via RMT methods.

\subsection{CG detection and optimal volumes for numerical simulations} 
In a real physical situation, we would like to know the amount of uncertainty we must take into acount due to our imperfect instruments. In other words, when an experiment is performed we would like to know the coarse graining we should consider in our measurements.\\
In the bipartite case, for example, this is easy to calculate from a set of experimental data using the least squares fitting method to compare the statistical analysis of the experimental results with the PDF given by \Eref{eq:PDF-2qubits}.

We confirmed this by performing a numerical Monte Carlo experiment with a sample of the same amount $n$ of data as in the previous simulations and a value of $p=1/2(1-h)$ chosen randomly. Using the method of least squares we recover an approximate value of $p_{\mathrm{fit}}$ whose difference with respect to the original $p$ will depend in the amount of data used for the fitting, and the choice of $\epsilon$ to define the volume of the neighborhood around the target state, in this case, the thickness of the spherical shell within $B^3$.

On the other hand, while performing our simulations, we would like to have a method to choose an optimal volume $V_{\epsilon}$ for estimating the probability density $P_N(p,\rts)$; if $\epsilon$ is too large, the numeric simulations will return only averaged values of the volumes for a large range of values of $\rts$, as expected, this leads to a PDF of an ensemble of random states which provides no information of the system. However, if $\epsilon$ is too small, the results of the simulations will contain considerably much noise and the data will be too much scattered away from the model it is suppose to fit. Thus, the choice of an optimal value of $\epsilon$ while performing our numerical simulations is another problem we would like to address, and to which we can approach with this same procedure of looking the best possible approximation for $p$.

\begin{center}
\begin{figure}[htbp]
\center
  \includegraphics[width=1\linewidth]{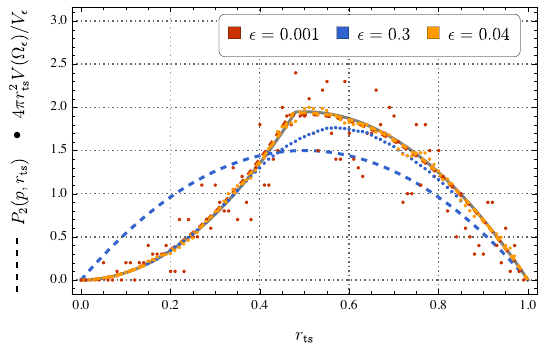}
  \caption{Approximation to $P_2(p,\rts)$ with $p=0.26$ (solid gray) using the least squares method (dashed lines), from the results of Monte Carlo simulations using  spherical shells of thickness $\epsilon$ in the Bloch ball, taking two extreme cases of $\epsilon= 0.001$ (red) and $\epsilon=0.3$ (blue) as described in the text, and one optimal value $\epsilon=0.04$. The numeric Monte Carlo simulations are displayed as scattered dots.}
  \label{fig:pfit_plot}
\end{figure}
\end{center}

For example, let's consider $p_{\mathrm{test}}=0.26$, and a sample of $n=10^4$ bipartite states that have been mapped to single qubit mixed states via CG map, then we may then use the least squares method to look for the value of $p$ that fits the best our data to the theoretical model $P_2(p,\rts)$, using two extreme cases as mentioned above $\epsilon=0.001,0.3,$ and $\epsilon=0.04$, obtaining $p_{\mathrm{fit}}(\epsilon)=0.2664, 0.4998, 0.2625$ respectively. With this results we can conclude that the optimal $\epsilon$ in this case is $\epsilon=0.04$ as it is shown in \Fref{fig:pfit_plot}.

Then, with this method is possible to detect not only the presence of CG in some experiment but also a very precise estimate of the amount of it, as for our numerical experiments it is possible to estimate an optimal value for the parameter $\epsilon$ that defines the volume of the neighborhood around the target state and with which we can calculate the volume of its preimage.
\section{Conclusions and outlook \label{sec:Conclusions}} 

In this paper, we have considered the problem of coarse graining in quantum many-body systems, focusing on the framework proposed in \cite{Pineda2021}, which models detectors with addressing errors that cannot perfectly resolve individual particles. Our analysis centered on two fundamental questions: (i) the characterization of the probability distribution over $N$-partite fine-grained states compatible with a coarse-grained observation, and (ii) the structure of the average state, particularly in the case where two particles are coarse-grained into a single subsystem.

The probability density function of a target state was obtained by computing the volume of the set of fine-grained states, associated with an infinitesimal neighborhood of the target state and distributed according to the appropriate Fubini-Study measure. 
This PDF also provides insight into the distribution of physical properties in quantum many-body systems. In fact, it identifies the most likely states under this hypothesis; a sharply peaked distribution suggests the emergence of potentially classical-like behavior. In the scenario under study, we derived this distribution analytically for the bipartite case using geometric arguments, and for larger systems using tools from random matrix theory. Both results are exact and analytical. For bipartite systems, 
the preimage volume of an infinitesimal neighborhood around the target state is constant for target states within a certain distance from the maximally mixed state, and vanishes at the surface of the Bloch sphere (i.e., for pure states). For larger systems, the probability density behaves like a Gaussian centered at the maximally mixed state, implying an exponential difficulty in observing pure states with imperfect measuring devices.

Regarding the average state, we obtained an exact analytical expression in the bipartite case. The resulting state exhibits, in general, quantum correlations. 

A particularly relevant finding is that the average preimage of the maximally mixed single-qubit state is not the maximally mixed two-qubit state, but rather an isotropic mixture with a singlet component. We also showed that, for a generic value of $h$, a pure target state can only arise from a coherent preimage state, which explains why the probability density vanishes in such cases.

Although the explicit computation of the average state was restricted to the case $N=2$, the unitary covariance of the coarse-graining map provides a pathway to generalize these results. This symmetry can be exploited to simplify the characterization of preimages in more complex systems. All analytical findings were supported by numerical simulations, which also serve as a practical tool for estimating the uncertainty introduced by measurement imperfections.

Important future research directions include analyzing preimage structures under more general coarse-graining schemes, such as mappings to multiple effective particles, or systems with local dimension $d > 2$. Using the structure of the average state to study the effective dynamics of coarse-grained systems constitutes an immediate next step. Furthermore, it would be highly desirable to explore connections between our approach and entropy-based reconstruction methods, as these could simplify the exploration of emergent dynamics under imperfect detection.
\section*{Acknowledgments} 
Useful discussions with Adán Castillo, Erick Navarrete, Raul Vallejo, Fernando de Melo and Alonso Botero are acknowledged. 
This work was supported by UNAM Posdoctoral Program (POSDOC), Estancias Posdoctorales por México-SECIHTI (formerly CONAHCYT), by UNAM-PAPIIT grants No. IG101324 (KU, CP) and IN112224 (CC, VRB, CIVM), SECIHTI projects CBF-2025-I-676 (CC, VRB, CIVM) and CBF-2025-I-1548 (KU, CP),  and UNAM PASPA–DGAPA.
\appendix
\section{\label{FSmeasure} Factorization of the Fubini–Study measure via a CG-adapted parametrization} 
For the two qubit case, a specific target state $\rho$ is represented by a vector $\mathbf{r}_{\rmt}$ in the Bloch sphere. From \Fref{fig:esferas} it can be observed that multiple states in $\mathbb{P}$ are mapped to the same target state $\rho$. Then, it is convenient to rewrite the FS measure in coordinates that are adapted to our problem of determining the inverse image of the CG map -- we use it later on in \Aref{Average-state}.  We will factorize the FS measure as the product of a measure defined on the preimage $\Omega_{\mathbf{R}}$ of a target state $\mathbf{R}$ in the Bloch sphere and a measure on the Bloch ball itself. 

\begin{figure}[h!]
    \centering
    \includegraphics[width=0.5\linewidth]{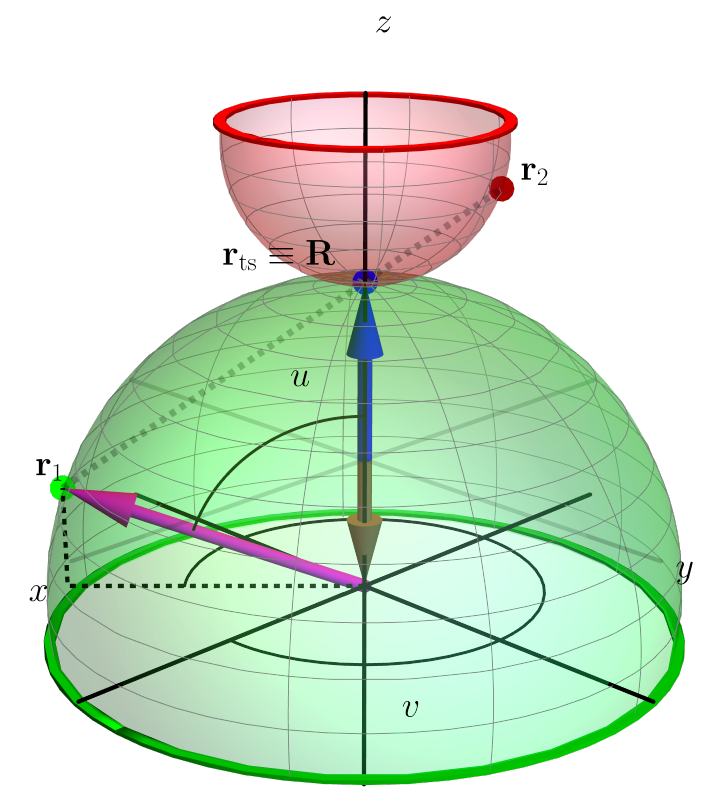}
    \caption{Subspheres corresponding to the reduced states that form a target state $\mathbf{R}$ aligned along the $z$-axis.
    The Bloch vectors $\mathbf{r}_1$ and $\mathbf{r}_2$ are shown as the green and red points, 
    respectively, while the vector of the target state $\mathbf{R}$ is represented in blue. 
    The vector $\mathbf{r}_1$ points in the direction of $\frac{1+h}{2h}\hat{\mathbf{m}}_{uv}$ (magenta arrow) 
    from the center of the green sphere $-\frac{1-h}{2h}\hat{\mathbf{k}}$ (brown arrow). 
    If $\mathbf{R}$ is fixed, once $\mathbf{r}_1$ is chosen, $\mathbf{r}_2$ is completely determined. 
    Hence, $(u,v)$ serve as coordinates in the preimage $\Omega_{\mathbf{R}}$.}
    \label{fig:appendixA_coords}
\end{figure}

As discussed in \Sref{sec:preimage_volume}, the Bloch vectors of the reduced states $\vb{r}_1$ and $\vb{r}_2$, lie on spheres of radii $R_1 = \rts(1+h)/2h$ and $R_2 = \rts(1-h)/2h$, with centers $\vb{c}_1 = -\vecrts(1-h)/2h$ and $\vb{c}_2 = \vecrts(1+h)/2h$, respectively. A convenient way to parametrize the reduced states is to place both spheres along the $z$-axis, and then rotate the $z$-axis onto the axis of a target state $\vb{R}$. This parametrization can be written as
\begin{align}
	&\vb{r}_i = R \mathcal{R}_{\Theta\Phi}\vb{a}_i(h,u,v), \nonumber\\ 
	&\vb{a}_1(h,u,v) = \frac{1}{2h}\qty( (1+h)\vu{m}_{uv}-(1-h)\vu{k}), \nonumber\\ 
	&\vb{a}_2 = \vb{a}_1(-h,u,v),
	\label{eq:r1r2 uv param}
\end{align}
where $\vu{m}_{uv}$ is a unit vector with angular coordinates $(u,v)$, which
parameterizes the sphere outside the origin where each of the vectors
$\vb{r}_i$ lies, see \Fref{fig:appendixA_coords}. The matrix $\mathcal{R}_{\Theta\Phi}$ represents a rotation by
an angle $\Theta$ along the axis $(-\sin\Phi,\cos\Phi,0)$, which maps the
$z$-axis to the $(\Theta,\Phi)$-position on the unit sphere along a geodesic. Notice that when $\Theta=0$, $\mathbf{r}_i = R\mathbf{a}_i$.

In this way, the Bloch vectors satisfy \Eref{eq:bipartite CGM} where $\vecrts\equiv\vb{R}$ has spherical
coordinates $(R,\Theta,\Phi)$. Now, we write the surface elements
$\dd\omega_i$ in \Eref{eq:vol element} as
\begin{equation}
	\dd\omega_i = r^{-3}\qty(x_i\dd{y_i}\dd{z_i}+\text{cyclic}),
\end{equation}
and using \Eref{eq:r1r2 uv param} we express them in terms of the variables $(u,v,R,\Theta,\Phi)$. This gives
\begin{equation}
	\dd\omega_1\dd\omega_2 = h^{-2}r^{-4}R^4\sin{u}\sin\Theta \dd{u}\dd{v}\dd\Theta\dd\Phi,
	\label{eq:d omega_1 d omega_2}
\end{equation}
with $r$ directly calculated from \Eref{eq:r1r2 uv param} as
\begin{equation}
	r^2 = r_1^2 = r_2^2 = \frac{R^2}{2h^2} (1+h^2-(1-h^2)\cos{u}).
	\label{eq:r and u}
\end{equation}
Furthermore, (\ref{eq:r and u}) can be rewritten as
\begin{equation}
	2h^2\kappa^2 = 1+h^2-(1-h^2)\cos{u}\, , 
	\label{eq:u and kappa}
\end{equation}
with $\kappa=r/R$ denoting the ratio between the radius of the reduced states and that of the target state. 
Differentiating the above expressions allows us to write the differentials $\dd{u}$ and $\dd{r}$ in \Eref{eq:d omega_1 d omega_2} and \Eref{eq:vol element}
in terms of the differentials $\dd\kappa$ and $\dd R$. Substituting these into \Eref{eq:vol element}, we obtain
\begin{equation}
	\dd\mu = \frac{3}{8\pi^3(1-h^2)} \dd\kappa\dd{v}\dd\gamma\dd{V} \equiv \dd{V_{\Omega}}\dd{V},
	\label{eq:dmu fact}
\end{equation}
with $\dd{V}=R^2\sin\Theta\dd{R}\dd\Theta\dd\Phi$ the Euclidean volume measure
on the Bloch sphere. The parameters $(R,\Theta,\Phi)$ are coordinates on the
CG map image, while $(\kappa,v,\gamma)$ are coordinates on the preimage
$\Omega_{\vb{R}}$ of a fixed target state $\vb{R}$, and $\dd V_\Omega$ is a
measure on such preimage. Since $\cos{u}\in[-1,1]$, $\kappa\in[1,h^{-1}]$, and
constrained by $r\leq 1$, we have $1\leq \kappa \leq \kappa_f$, with $\kappa_f
= h^{-1}$ if $R\leq h$, and $\kappa_f = R^{-1}$ if $h<R\leq 1$. The coordinates
$v$ and $\gamma$ range from $0$ to $2\pi$.

In the case of separable states, all states are of the form $\ket{\Psi}=\ket{\vb{n}_1}\ket{\vb{n}_2}$, so $(\theta_1,\phi_1,\theta_2,\phi_2)$ are coordinates in this space. The Fubini-Study metric in the bipartite state space induces an invariant metric in the space of separable states with measure $\dd\mu^{\otimes} = \dd\omega_1 \dd\omega_2/16\pi^2$, where $16\pi^2$ is the total volume of separable states. Taking into account that separable states satisfy $r=1$, by differentiating \Eref{eq:r and u} we can express $\dd{u}$ in terms of $\dd{R}$ and substitute it into equation (\ref{eq:d omega_1 d omega_2}) to obtain
\begin{equation}
	\dd\mu^{\otimes} = \frac{\dd{v}\dd{V}}{4\pi^2(1-h^2)R},
	\label{eq:dmu fact ss}
\end{equation}
where the range of $\kappa$ now imposes that $h\leq R\leq 1$, while the other variables span their entire range.
\section{\label{sec:Volumen} Volume integrals for two-qubit system} 
We calculate the volume $V(\Omega_\epsilon)$ of the preimage of a neighborhood
$V_\epsilon$ in the Bloch ball by integrating the measure in (\ref{eq:dmu
fact}), first over the preimage $\Omega_{\vb{R}}$ of each $\vb{R}$ in
$V_\epsilon$, and then over the entire $V_\epsilon$. For this, we consider
that, in the most general case, there is a region $V_{in}$ in $V_\epsilon$
within a sphere of radius $h$, where $\kappa\in\qty[1,h^{-1}]$, and another
region $V_{out}$ outside the sphere of radius $h$, where
$\kappa\in\qty[1,R^{-1}]$. This results in
\begin{align}
	V(\Omega_\epsilon) 
	&=
	 \int_{V_\epsilon}\dd{V} \int_{\Omega_{\vb{R}} }\dd{V_\Omega} 
	 \nonumber
	 \\
	 &=
	 \frac{3}{2\pi h(1+h)}\int_{V_{\text{in} }} \dd{V} 
	 \nonumber
	 \\
	& \quad + \frac{3}{2\pi(1-h^2)}\int_{V_{\text{out} }} \dd{V} (R^{-1} - 1)
	 \, .
\label{eq:vol}
\end{align}
Expression (\ref{eq:vol}) is not subject to any approximation and can be integrated numerically for any $V_\epsilon$ within the Bloch ball. However, we will associate the neighborhood $V_\epsilon$ with the experimental error bars of the measurements determining the state $\rho $. As a first approximation, we will assume that these error bars define an infinitesimal neighborhood centered at the Bloch vector $\vecrts$. In this case, we can disregard the transition in which $V_\epsilon$ changes from being entirely within the sphere of radius $h$ to being entirely outside, as only these two cases are relevant. We will also take the integrands as constants, equal to their value at $\vecrts$. The volume then reduces to 
\begin{equation}
	V(\Omega_\epsilon(\rts)) = \frac{3 V_\epsilon}{2\pi(1+h)}
    \begin{cases}
    \frac{1}{h} & \text{if } 0 \leq \rts < h, \\[5pt]
    \frac{1-\rts}{(1-h) \rts} & \text{if } h < \rts < 1,
    \end{cases}
\label{eq:Vol aprox}
\end{equation}
where $\Omega_\epsilon(\rts)$ is the preimage of an infinitesimal neighborhood $V_\epsilon$ at a distance $\rts$ from the origin. From now on, we will use $V_\epsilon$ to refer both to the neighborhood and its volume.

It is important to mention that expression (\ref{eq:Vol aprox}) does not depend on the shape of the neighborhood $V_\epsilon$ but only on its volume. Additionally, this expression is not valid for a neighborhood centered at the origin ($\rts=0$) when $h\rightarrow 0$, as it is not possible to evaluate this limit with $V_\epsilon$ entirely inside or entirely outside the sphere of radius $h$. To obtain a finite value in this limit, we can analytically calculate expression (\ref{eq:vol}) for a $V_\epsilon$ as a ball of radius $\epsilon$ centered at the origin, with $h<\epsilon$, where $V_{in}$ is defined by $0\leq R \leq h$ and $V_{out}$ by $h< R\leq\epsilon$. This results in
\begin{equation}
   V(\Omega_\epsilon(0)) = \frac{\epsilon^2(3-2\epsilon) - h^2}{1-h^2}, \quad h<\epsilon,
\end{equation}
which, in the limit as $h\rightarrow 0$, results in $\epsilon^2(3-2\epsilon)$.

For separable states, the neighborhood $V_\epsilon$ can only be outside the sphere of radius $h$. By directly integrating the measure (\ref{eq:dmu fact ss}), we obtain
\begin{equation}
	V(\Omega_\epsilon^\otimes) = \frac{1}{2\pi(1-h^2)}\int_{V_\epsilon}\frac{\dd{V}}{R}.
\label{eq:Vol ss}
\end{equation}
Again, taking the approximation of $V_\epsilon$ as infinitesimal at a distance $\rts$ from the origin, we find that
\begin{equation}
	V(\Omega_\epsilon^\otimes(\rts)) = \frac{V_\epsilon}{2\pi(1-h^2)\rts}, \quad h<\rts<1.
\label{eq:Vol ss epsilon}
\end{equation}

\section{\label{RMT-Method-N} RMT methods for $N$ qubits} 
This appendix aims to present the generalization of the methods implemented in
the previous appendix for application in systems composed of $N$ qubits. To
begin, we introduce the notation that allows us to efficiently express the
$N$-partite state and the diagonal elements of the target state. Subsequently,
we represent the joint probability density function of the diagonal elements of
the target state in integral form and solve these integrals in the Laplace
space. We apply the Vandermonde operator of partial derivatives and invert the
Laplace transform using the residue theorem. As a final result, we obtain the
probability density function of the target state derived from states composed
of $N$ qubits, as a function of the Bloch radius of the target state and the
set of probabilities that define the CG map.
We once again note that the target state considered is always monopartite, 
hence $D=2$.\par
We start with \Eref{eq:appF-Derivative-principle} which states a relationship
between $f_{\rho}(\vec{\lambda})$, the joint PDF of the eigenvalues of $\rho$
and $\left.\Psi_{\rho}(\vec{\rho})\right|_{\vec{\rho}=\vec{\lambda}}$, the
joint PDF of its diagonal elements (evaluated in the eigenvalues of $\rho$).

We consider a pure $N$-partite state of the form
\begin{align}
	\ket{\psi} = \sum_{l }c_l \ket{l}, \qquad 
	l=(l_1,\ldots,l_{N}) \label{eq:appG-N-state}
\end{align} 
where $l_i=0,1$ represents the state of the $i$-th qubit, and $\ket{l}$ is
an element of the computational basis. \\
Density matrices of $N$ qubits are distributed according the FS measure \cite{Zyczkowski1999}, their diagonal elements $x_l=|c_l|^2$ so that $\sum_{l}x_{l}= 1$ and $x_{l}\geq 0$, have a uniform distribution over the $(2^N-1)$-simplex.\\
We may then exclude the last element ${l=(1,\dots,1)}$ and implement the condition $0\leq 1-\sum_{l\neq(1,\dots,1)} x_l$ so we can write the distribution of the diagonal elements of $\rho$ as
\bq
\dd\nu = (2^N-1)! \prod_{l\neq(1,\dots,1)}\dd x_l,
\eq 
where the constant factor correspond to the volume of the standard $(2^N-1)$-simplex.\\
Applying the CG map on the state (\ref{eq:appG-N-state}) we obtain a target state whose diagonal elements are of the form
\bq
	\rho_{00} = \sum_{l} \overline{l}\cdot p \ x_{l}, \qquad
	\rho_{11} = \sum_{l} l\cdot p \ x_{l},
	\label{appG:diagonal-N-elements-target-1} 
\eq
where $\overline{l} = (\overline{l}_{1}, \ldots \overline{l}_{N})$ with $\overline{l}_{i} = 1-l_{i}$ and $p=(p_{1}, \ldots, p_{N})$.
As in the case when $N=2$, it must hold that $\sum_i p_i =1$, which implies that $\rho_{00}+\rho_{11}=1$, defining thus a mapping from the $(2^N-1)$-simplex to the 1-simplex. The joint probability density function of the diagonal elements $(\rho_{00},\rho_{11})$ of $\rho$, is now given only in terms of $\rho_{00}$ by the integral
\begin{align}
	\Psi(\rho_{00}) &= (2^N-1)!\int_H \prod_{l\neq(1,\dots,1)} \dd x_l \notag \\  
	&= (2^N-1)!\int_{\mathbb{R}^{S}_{+}} \prod_{\ell}\!\dd x_\ell\;\Theta(\rho_{00}-\tilde{\rho}_{00})\Theta(\rho_{11}-\tilde{\rho}_{11}) \notag \\ 
	&\equiv (2^N-1)! A(\rho_{00},\rho_{11})
\end{align}  
where $S=2^N-2$, $\ell$ index the elements of the computational basis excluding the first $(0,\dots,0)$ and last $(1,\dots,1)$ elements. $H$ represents the preimage of $\rho_{00}$, defined as the region where the inequalities $0\leq x_\ell$, ${0\leq \rho_{00}-\tilde{\rho}_{00} = x_{0\dots0}}$, and ${0\leq \rho_{11}-\tilde{\rho}_{11} = x_{1\dots1}}$ hold, with  
\begin{align} 
	\tilde{\rho}_{00} &= \sum_{\ell} \overline{\ell}\cdot p \ x_{\ell}, 
	               \label{appF:generalized_diagonal-elements-target-0-tilde} \\  
	\tilde{\rho}_{11} &= \sum_{\ell} \ell \cdot p \ x_{\ell}. 
	               \label{appF:generalized_diagonal-elements-target-1-tilde}
\end{align}
Now, in order to calculate the operator
$\Delta(\vec{\rho})\Psi_{\rho}(\rho_{00},\rho_{11})$, with
$\vec{\rho}=(\rho_{00},\rho_{11})$, we will calculate the Laplace transform of the
area function $A(\rho_{00},\rho_{11})$. Let us proceed,
\begin{align}
\tilde{A}(s_0,s_1)&=\mathcal{L}_{s_0,s_1}\left[A(\rho_{00},\rho_{11})\right]\notag \\
 &=\int_{\mathbb{R}^{S}_{+}} \prod_{\ell}\!\dd x_\ell\;\mathcal{L}_{s_0,s_1}\left[\Theta(\rho_{00}-\tilde{\rho}_{00})\Theta(\rho_{11}-\tilde{\rho}_{11}) \right]\notag\\
 &=\frac{1}{s_0s_1}\int_{\mathbb{R}^{S}_{+}} \prod_{\ell}\!\dd x_\ell\;e^{-s_0\tilde{\rho}_{00}}e^{-s_1\tilde{\rho}_{11}}\notag\\
 &=\frac{1}{s_0s_1}\prod_{\ell}\int_{\mathbb{R}^{S}_{+}} \!\dd x_\ell\;e^{-(\bar{\ell}\cdot p\,s_0+\ell\cdot p\, s_1)x_\ell}\notag\\
 &=\prod_l\;(\bar{l}\cdot p\,s_0+l\cdot p\, s_1)^{-1},
 \label{appG:Prob-Laplace-space-N}
\end{align}
the change in the index $\ell$ to $l$ in the last step of the previous calculation comes from the result that $\sum_i p_i =1$, thus, when $l=(1,\dots,1)$ we obtain $l\cdot p\, s_i = s_i$ and $\bar{l}\cdot p\, s_i = 0$ , and analogously for $l=(1,\dots,1)$.

Then, applying the operator $\Delta(-\partial_{(\rho_{00},\rho_{11})})$ in the Laplace space and subsequently inverting the transform, we get
\begin{widetext}
    \begin{align}
	\Delta\!\left(-\partial_{\vec{\rho}}\right)A(\rho_{00},\rho_{11}) &= \mathcal{L}^{-1}_{\rho_{00},\rho_{11}} \left[(s_{0}-s_{1})\tilde{A}(s_{0},s_{1})\right] \notag \\
	&= \frac{1}{(2\pi i)^2}\oint_{\gamma}\oint_{\gamma}\dd s_0 \dd s_1\; (s_0-s_1)\prod_l\;(\bar{l}\cdot p\,s_0+l\cdot p\, s_1)^{-1}e^{s_0 \rho_{00}}e^{s_1 \rho_{11}}\notag\\
	&= (2^{N}-1)! \sum \text{Res} \Big[(s_{0}-s_{1}) \times \prod_{l}(\bar{l}\cdot p\,s_0+l\cdot p\, s_1)^{-1}e^{s_0 \rho_{00}}e^{s_1 \rho_{11}} ;s_{0},s_{1} \Big],
\end{align}
\end{widetext}
where $\gamma$ is a simple closed curve sufficiently large so it contains all the poles of the function.

We begin by calculating the residues with respect to $s_{0}$, which arise when
$s_{0}=-\frac{l'\cdot p}{\overline{l'}\cdot p}s_{1}$, excluding the case 
$\overline{l'}=(0,\dots,0)$. When, $l\cdot p$ yields a distinct value for each $l$, there are $2^{N}-1$ simple poles. Consequently, for a given $l'$, the residue is:
\begin{widetext}
\begin{align} 
	& \lim\limits_{s_{0} \rightarrow -\frac{l' \cdot p}{\overline{l'}\cdot p} s_{1}}\left[
	\left(s_{0}+\frac{l'\cdot p}{\overline{l'} \cdot p}s_{1} \right)(s_{0}-s_{1} ) \prod_{l}(\overline{l}\cdot p \, s_{0} + l\cdot p\, s_{1})^{-1}
	e^{s_{0}\rho_{00}}e^{s_{1}\rho_{11}}  \right] \notag \\
	& \hspace{2cm} =  -  \prod_{\underset{l\neq l'}{l}}\left\lbrace\left[(l\cdot p)(\overline{l'}\cdot p)-(\overline{l}\cdot p)(l'\cdot p)\right]^{-1}(\overline{l'}\cdot p)^{2^{N}-3}s_{1}^{-(2^{N}-2)} e^{s_{1}\left(a_{1}-\frac{l'\cdot p}{\overline{l'}\cdot p}\rho_{00} \right)} \Theta\!\left(a_{1}-\frac{l'\cdot p}{\overline{l'}\cdot p}\rho_{00} \right) \right\rbrace. \label{eq:polo-s0}
\end{align}
\end{widetext}  
The expression (\ref{eq:polo-s0}) has only one pole with respect to $s_{1}$, which is a pole at zero of order $2^{N}-2$. We use the Laurent expansion
\bq 
	\frac{e^{s_{1}\left(\rho_{11}-\frac{l'\cdot p}{\overline{l'}\cdot p}\rho_{00} \right)}}{s_{1}^{(2^{N}-2)}}
	= \sum_{n=0}^{\infty} \frac{\left(\rho_{11}-\frac{l'\cdot p}{\overline{l'}\cdot p}\rho_{00} \right)^{n}}{n!}s_{1}^{n-(2^{N}-2)}
\eq
to find that the residue takes the form
\bq 
	\frac{1}{(2^{N}-3)!}\left(\rho_{11}-\frac{l'\cdot p}{\overline{l'}\cdot p}\rho_{00} \right)^{2^{N}-3}.
\eq 
Finally, we can compile the previous results by summing all the residues to determine that
\begin{align}
	&\Delta\!\left(-\partial_{\vec{\rho}^{{}\,\rmt}}\right)\Psi(\rho_{00},\rho_{11}) \notag \\
	& \qquad = - \frac{(2^{N}-1)!}{(2^{N}-3)!} \sum_{l'\neq(0,\dots,0)}\left(\overline{l'}\cdot p\, \rho_{11} - l'\cdot p\,\rho_{00}\right)^{2^{N}-3} \notag \\
	& \qquad \quad \times \Theta\!\left(\overline{l'}\cdot p\,\rho_{11}-l'\cdot p\,\rho_{00} \right) \\
	& \qquad \quad \times \prod_{l\neq l'}\left[(l\cdot p)(\overline{l'}\cdot p)-(l'\cdot p)(\overline{l}\cdot p) \right]^{-1}. \notag 
\end{align}
\par 
We can finally gather the result in a compact expression. We define a transformed
vector, that depends both from the set of probabilities $p$ and the multi-index $l$,
with components $\tilde p_i^{(l)} = (2l_i -1) p_i$. $\tilde p^{(l)}$ is the probability vector
with signs flipped according to $l$. We end up with 
\begin{multline}
P_N(p,\rts) = c_N\; \rts \\
\times \sum_{l\neq(0,\dots,0)}\frac{ \left( \Sigma_i \tilde p_i^{(l)} - \rts \right)^{2^ {N}-3}
\Theta\!\left( \Sigma_i \tilde p_i^{(l)} - \rts\right) }
{\prod\limits_{l'\neq(0,\dots,0)} l'\cdot \tilde p^{(l)} } 
\label{appG:Prob-Vol-Gen-N}
\end{multline}
with the normalization constant $c_N=(2^{N}-1)(2^{N}-2)/2^{2^{N}-2}$.
Notices that due to the unitary invariance of the problem, $\rts$ can be taken
as the radius of the Block vector independently of the direction of the polarization
of the state.
The above formula constitutes the main result of the appendix. 

Note that the analytical formula in \Eref{appG:Prob-Vol-Gen-N} does not admit
cases where some pair of probabilities $p_k$ are equal. Such cases can be dealt
with before inverting the Laplace transforms. 
For example, the case with all probabilities being equal, $p_{k}=\frac{1}{N}$,
leads to the formula
\begin{align}
	\tilde{\Psi}\left(s_{0},s_{1}\right) =
	N^{2^{N}}(2^{N}-1)!
	\prod_{n=0}^{N} \left[s_{0}(N-n) +s_{1}n\right]^{-\binom{N}{n}}.
	\label{appG:PDFmarginalGenEqui}
\end{align}
In this case, the poles of the function to be inverted are of binomial order
as a function of $N$, which makes it difficult to find an analytical expression
for arbitrary $N$. However, once the size of the system is fixed, 
it is possible to apply Laplace transform techniques to invert
the corresponding expressions.
\section{\label{Average-state} Average state integrals} 
We define the average state in the preimage of a neighborhood $V_\epsilon$ as 
\begin{equation}
	\varrho^{\avs} = \frac{1}{V(\Omega_\epsilon)}\int_{\Omega_\epsilon} \dd\mu \varrho,
	\label{eq:def AS}
\end{equation}
where $\varrho$ is a state in the space of bipartite pure states. Since the trace is linear, it is straightforward to see that $\tr{\varrho^{\avs}}=1$. Additionally, since the measure $\dd\mu$ is non-negative, the integral divided by the volume $V(\Omega_\epsilon)$ represents a convex sum of pure states $\varrho$, which generally results in a mixed state, not a pure state.

From \Eref{eq:app-parametrized-state}, it follows that a density matrix in the space of pure states takes the form
\begin{align}
	\varrho 
	&=
	\cos^2\frac{\eta}{2}\ketbra{\vb{n}_1}\otimes\ketbra{\vb{n}_2}  
	\nonumber
	\\ 
	& \quad {}+\sin^2\frac{\eta}{2}\ketbra{-\vb{n}_1}\otimes\ketbra{-\vb{n}_2}  
	\nonumber
	\\ 
	& \quad {}+\frac{1}{2}\sin\eta\qty(e^{\ii\gamma}\ketbra{-\vb{n}_1}{\vb{n}_1}\otimes\ketbra{-\vb{n}_2}{\vb{n}_2} + \text{h.c.} ),
\end{align}
where the third term is cyclic in $\gamma$, so it does not contribute to the integral with the measure in \Eref{eq:dmu fact}. Taking the density matrices $\ketbra{\vb{n}_i}=\frac{1}{2}(I+\vb{n}_i\cdot\vb*\sigma)$ and expanding the tensor products, the integral in (\ref{eq:def AS}) becomes
\begin{align}
	\int \dd\mu \varrho 
	&=
	 \frac{1}{4}\int \dd\mu 
	 \left( \rule{0ex}{3ex} \right.
	 I \otimes I + (\vb{r}_1\cdot\vb*{\sigma})\otimes I 
	 \nonumber
	 \\
	 &
	\quad
	{}+ I\otimes (\vb{r}_2\cdot\vb*{\sigma}) 
	+ \sum_{ij}n_{1i}n_{2j}
	\sigma_i\otimes\sigma_j 
	\left.
	\rule{0ex}{3ex}
	\right),
	\label{eq:dmu rho}
\end{align}
where we recall that $\vb{r}_i=r \, \vb{n}_i$ with $r=\cos\eta$. In this way,
the problem reduces to integrating the Bloch vectors $\vb{r}_i$ and the
components of the tensor product $\vb{n}_1\otimes\vb{n}_2$ on
$\Omega_\epsilon$. We use the parameterization in \Eref{eq:r1r2 uv param} of
the Bloch vectors, together with \Eref{eq:u and kappa} to calculate
the integrals first over the preimage  $\Omega_{\vb{R}}$ in $(\kappa,v,\gamma)$
coordinates and then over all $\vb{R}$ in the neighborhood $V_\epsilon$.

The integrals of the Bloch vectors can be expressed as
\begin{equation}
	\int_{\Omega_\epsilon} \dd\mu \vb{r}_i = \int_{V_\epsilon} \dd{V} R \mathcal{R}_{\Theta\Phi} \int_{\Omega_{\vb{R} }} \dd{V_\Omega}\vb{a}_i,
\end{equation}
where, by direct integration, it follows that
\begin{equation}
	\int_{\Omega_{\vb{R} }} \dd{V_\Omega} \vb{a}_1 = \frac{ (\kappa_f-1)(3-h (\kappa_f^2+\kappa_f+1)) }{2\pi(1-h)^2(1+h)}\vu{k},
\end{equation}
and that the integral of $\vb{a}_2$ is obtained by exchanging $h$ for $-h$. By performing the integration over $V_\epsilon$, we obtain
\begin{align}
	\int_{\Omega_\epsilon} \dd\mu \,  \vb{r}_1 
	&= 
	-\frac{1-h}{2\pi h^2(1+h)} 
	A
	+ \frac{1}{2\pi(1-h^2)(1-h)}
	B
	\, , 
	\nonumber 
	\\ 
	\int_{\Omega_\epsilon} \dd\mu \vb{r}_2 
	&= 
	\frac{1+4h+h^2}{2\pi h^2(1+h)^2}
	A
	+ \frac{1}{2\pi(1-h^2)(1+h)}
	C
	\, ,
	\label{eq:ri integrals}
\end{align}
with $V_{in}$ and $V_{out}$ defined as in \Eref{eq:vol}, and
\begin{align*}
A
&=
\int_{V_{in}} \dd{V}\vb{R}\, ,
\\
B
&=
\int_{V_{out}} \dd{V} (R^{-1}-1)(3-h(R^{-2}+R^{-1}+1))\vb{R}\, ,
\\
C
&=
\int_{V_{out}} \dd{V}(R^{-1}-1)(3+h(R^{-2}+R^{-1}+1))\vb{R}\, .
\end{align*}

We compute the integral of $n_{1i}n_{2j}$ as the components of the integral of the tensor product as follows
\begin{align}
	\int_{\Omega_\epsilon} \dd\mu \vb{n}_1\otimes\vb{n}_2 
	&=
	 \int_{\Omega_\epsilon} \dd\mu (\kappa R)^{-2}\vb{r}_1\otimes\vb{r}_2 
	 \nonumber
	 \\
	 &=
	  \int_{V_\epsilon}\dd{V} 
	  \qty(
	  \mathcal{R}_{\Theta\Phi}\otimes\mathcal{R}_{\Theta\Phi}) \\
	  & \qquad \times \int_{\Omega_{\vb{R} }}\dd{V_\Omega} 
	  \kappa^{-2}\vb{a}_1\otimes\vb{a}_2
	  \, .
\end{align}
A direct integration yields
\begin{align}
	\int_{\Omega_{\vb{R} }} \dd{V_\Omega} \kappa^{-2} \vb{a}_1\otimes\vb{a}_2 
	&=
	 \frac{ (\kappa_f-1)^2 \qty(h^2 \kappa_f (\kappa_f+2)-3)}{4\pi\qty(1-h^2)^2 \kappa_f} 
	 \mathbb{I}
	 \nonumber
	 \\
	 & 
	 {}\quad
	 - \frac{3(\kappa_f^2-1) \qty(h^2 \kappa_f^2-1)}{4\pi\qty(1-h^2)^2 \kappa_f} \vu{k}\otimes\vu{k}
	 \, ,
\end{align}
where $\mathbb{I}=\vu{i}\otimes\vu{i}+\vu{j}\otimes\vu{j}+\vu{k}\otimes\vu{k}$.
Using that $\mathbb{I}$ is invariant under the product of rotations $\qty(\mathcal{R}_{\Theta\Phi}\otimes\mathcal{R}_{\Theta\Phi})$, we have
\begin{align}
	\int_{\Omega_\epsilon} \dd\mu \, \vb{n}_1\otimes\vb{n}_2 
	&=
	 \left(
	 -\frac{(1-h)A'}{2\pi h(1+h)^2}
	 +
	 \frac{B'}{4\pi\qty(1-h^2)^2}
	 \right) \mathbb{I}
	 \nonumber
	 \\
	& \qquad
	{} - \frac{3C'}{4\pi\qty(1-h^2)^2}
	 \, ,
	\label{eq:n1n2 integral}
\end{align}
where
\begin{align*}
A'
&=
\int_{V_{in}}\dd{V} \, ,
\\
B'
&=
\int_{V_{out}}\dd{V} R(R^{-1}-1)^2 \qty(h^2 R^{-1}(R^{-1}+2)-3)\, ,
\\
C'
&=
 \int_{V_{out}}\dd{V}R^{-1}(R^{-2}-1) \qty(h^2 R^{-2}-1) \vb{R}\otimes\vb{R}
 \, .
 \end{align*}
Now, we take $V_\epsilon$ as an infinitesimal neighborhood at a distance $\rts$ from the origin. The integrals (\ref{eq:ri integrals}) and (\ref{eq:n1n2 integral}) separate into two expressions: when the neighborhood is entirely within the sphere of radius $h$, and another when it is entirely outside. By substituting these expressions into \Eref{eq:def AS}, along with the volume $V(\Omega_\epsilon(\rts))$ from \Eref{eq:Vol aprox}, we finally obtain
\begin{align}
	\varrho^{\avs} 
	&=
	 \frac{1}{4}I\otimes I + c_1 (\vu{r}_{\rmt}\cdot\vb*\sigma)\otimes I + c_2 I\otimes (\vu{r}_{\rmt}\cdot\vb*\sigma)
	 \nonumber
	 \\
	 &
	 \quad
	 {}+ c_3(\vu{r}_{\rmt}\cdot\vb*\sigma)\otimes (\vu{r}_{\rmt}\cdot\vb*\sigma) + c_4 \sum_i\sigma_i\otimes \sigma_i
	 \, ,
	\label{eq:varrho}
\end{align}
with the following $c_i$ coefficients
\begin{widetext}
\begin{equation}
	\mqty(c_1 \\c_2 \\c_3 \\c_4) = 
    \begin{cases}
    \frac{1}{12h(1+h)}
    \mqty(
    -(1-h^2)\rts \\ 
    (1+4h+h^2)\rts \\ 
    0 \\ 
    -h(1-h)
    ) & \text{if } 0 \leq \rts < h, \\[30pt]
    \frac{1}{24(1-h^2)\rts^2} 
    \mqty( 
    2\rts(1+h)( 3\rts^2-h(\rts^2+\rts+1) ) \\ 
    2\rts(1-h)( 3\rts^2+h(\rts^2+\rts+1) ) \\ 
    3(1+\rts)(\rts^2-h^2) \\ 
    -(1-\rts)(3\rts^2-h^2(2\rts+1))
    ) & \text{if } h < \rts < 1.
    \end{cases}
\label{eq:ci aprox}
\end{equation}
\end{widetext}
In the space of separable bipartite states, we define the average state of the preimage of a neighborhood $V_\epsilon$ analogously as
\begin{equation}
	\varrho^{\avs} = \frac{1}{V(\Omega^{\otimes}_\epsilon)}\int_{\Omega^{\otimes}_\epsilon} \dd\mu^{\otimes} \varrho,
	\label{eq:defASss}
\end{equation}
where all density matrices of separable states take the form 
\begin{align}
	\varrho 
	&= 
	\ketbra{\vb{n}_1}\otimes\ketbra{\vb{n}_2} 
	\nonumber
	\\
	&= 
	\frac{1}{4}\Big( I\otimes I+(\vb{n}_1\cdot\vb*{\sigma})\otimes I
	\nonumber
	\\
	& 
	\quad
	{}+ I\otimes (\vb{n}_1\cdot\vb*{\sigma})+\sum_{ij} n_{1i}n_{2j}\sigma_i \otimes \sigma_j \Big),
\end{align} 
reducing the problem to integrating the vectors $\vb{n}_i$ and the tensor product $\vb{n}_1\otimes\vb{n}_2$ similarly to the general case. We use the parameterization given in \Eref{eq:r1r2 uv param}, and together with \Eref{eq:r and u} and $r=1$, we express $u$ in terms of $R$ to perform the integration with the measure $\dd\mu^{\otimes}$ in \Eref{eq:dmu fact ss}. All integrals in the variable $v$ are straightforward, yielding
\begin{equation}
	\int_{\Omega^{\otimes}_\epsilon} \dd\mu^{\otimes}\vb{n}_1 = \frac{1}{2\pi(1-h^2)(1-h)} \int_{V_\epsilon}\dd{V} \frac{R^2-h}{R^3}\vb{R},
\end{equation}
the integral of $\vb{n}_2$ is obtained from the previous one by exchanging $h$ with $-h$, and
\begin{equation}
	\int_{\Omega^{\otimes}_\epsilon} \dd\mu^{\otimes}\vb{n}_1\otimes\vb{n}_2 
	=
		\frac{1}{4\pi(1-h^2)^2} \int_{V_\epsilon} \dd{V} 
		\left( A''-B'' \mathbb{I} \right)
		\, ,
\end{equation}
where
\begin{align*}
A''
&=
\int_{V_\epsilon} \dd{V} 
		R^{-5}(R^4+R^2+h^2(R^2-3)) \vb{R}\otimes\vb{R}
\, ,
\\
B''
&=
\int_{V_\epsilon} \dd{V} 
R^{-3}(1-R^2)(R^2-h^2) 
\, .
\end{align*}
Recall that the preimage of a target state contains separable states only if it is outside the sphere of radius $h$ in the Bloch ball, for this reason, there is a unique integration region unlike the general case. Taking $V_\epsilon$ as an infinitesimal neighborhood at a distance $\rts>h$ from the origin, and using the volume computed in \Eref{eq:Vol ss epsilon}, we find that the average state has the same form as in \Eref{eq:varrho} but with the following coefficients
\begin{align}
	c_1 &= \frac{\rts^2-h}{4(1-h)\rts} \nonumber\\ 
	c_2 &= \frac{\rts^2+h}{4(1+h)\rts} \nonumber\\ 
	c_3 &= \frac{\rts^4+\rts^2+h^2(\rts^2-3)}{8(1-h^2)\rts^2} \nonumber\\
	c_4 &= -\frac{(1-\rts^2)(\rts^2-h^2)}{8(1-h^2)\rts^2}.
	\label{eq:ci aprox ss}
\end{align}

\begin{figure}[htbp]
\centering
  \includegraphics[width=1\linewidth]{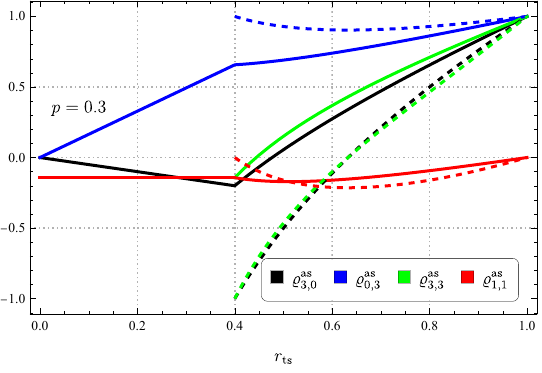}
  \caption{Components of the average state  $\varrho^{\avs}$ as a function of the $\rts$. The components are in the tensor product of the Pauli basis. The graphs with solid lines shows the case where the set of pure bipartite states over which it was averaged includes both entangled and non-entangled states.The graphs with dotted lines represent the case in which the averaging was performed over a set of pure bipartite states that are non-entangled, i.e., are product states.}
  \label{fig:avg_state_rhos}
\end{figure}

The general expression in \Eref{eq:varrho} for the average state explicitly reveals that the functions $c_i$ are independent of the polarization of $\rho$. Furthermore, by reorienting $\vb{r}_{\rmt}$, these functions can be determined as averages of expectation values. If we set $\vu{r}_{\rmt}=\vu{k}$, the average state takes the form 
\begin{equation}
    \varrho^{\avs} = \frac{1}{4}\sigma_{0,0} + c_1\sigma_{3,0} + c_2\sigma_{0,3} + c_3\sigma_{3,3} + c_4\sum_{i=1}^{3}\sigma_{i,i},
    \label{eq:avg state Pauli expansion}
\end{equation}
where $\sigma_{\mu_1,\mu_2}$ is defined as in \Eref{varrhoexpands}. It is then straightforward to observe that the expectation values averaged in \Eref{eq:varrho_nu} are expressed in terms of the functions $c_i$ as follows:  
\begin{align}
    \varrho_{0,0}^{\avs} &= 1, \nonumber \\
    \varrho_{3,0}^{\avs} &= 4 c_1, \nonumber  \\
    \varrho_{0,3}^{\avs} &= 4 c_2, \nonumber \\
    \varrho_{3,3}^{\avs} &= 4(c_3 + c_4), \nonumber \\
    \varrho_{1,1}^{\avs} &= \varrho_{2,2}^{\avs} = 4c_4,  
\end{align}
with $\varrho_{\mu_1,\mu_2}^{\avs} = 0$ otherwise.  
That is, in the basis of Hermitian matrices $\sigma_{\mu_1 \mu_2}$, the average state $\varrho$ for a target state polarized along the $z$-axis has only six nonzero entries, four of which are independent. These can be measured as averages of expectation values and are directly comparable to the functions $c_i$ derived here. \Fref{fig:avg_state_rhos} displays the coefficients $\varrho_{\mu_1 \mu_2}^{\avs}$  for both the general case and the separable case with $N=2$.

\bibliography{apssamp}

@PREAMBLE{
 "\providecommand{\noopsort}[1]{}" 
 # "\providecommand{\singleletter}[1]{#1}%" 
}

@article{firstconcurrence,
	Author = {S. Hill and W. K. Wootters},
	Journal = prl,
	Number = {26},
	Pages = {5022},
	Publisher = {APS},
	Title = {Entanglement of a Pair of Quantum Bits},
	Volume = {78},
	Year = {1997}}

@book{Bengtsson2008,
address = {Cambridge},
author = {Bengtsson, I. and {\.{Z}}yczkowski, K.},
pages = {107},
publisher = {Cambridge University Press},
title = {{Geometry of quantum states: An Introduction to Quantum Entanglement}},
year = {2008}
}

@article{Christandl2014,
author = {Christandl, M. and Doran, B. and Kousidis, S. and Walter, M.},
journal = {Commun. Math. Phys.},
pages = {1-52},
title = {{Eigenvalue distributions of reduced density matrices}},
volume = {332},
year = {2014}
}

@article{Mejia2017,
author = {Mej{\'{i}}a, J. and Zapata, C. and Botero, A.},
journal = {J. Phys. A: Math. Theor.},
pages = {025301},
title = {{The difference between two random mixed quantum states: exact and asymptotic spectral analysis}},
volume = {50},
year = {2017}
}

@article{Kieburg2023,
author = {Kieburg, M. and Zhang, J.},
journal = {Adv. Math.},
pages = {108833},
title = {{Derivative principles for invariant ensembles}},
volume = {413},
year = {2023}
}

@article{Pineda2021,
author = {Pineda, Carlos and D{\'{a}}valos, David and Viviescas, Carlos and Rosado, Antonio},
journal = {Phys. Rev. A},
pages = {042218},
title = {{Fuzzy measurements and coarse graining in quantum many-body systems}},
volume = {104},
year = {2021}
}

@article{Ekert1995,
author = {Ekert, A. and Knight, P. L.},
journal = {Am. J. Phys.},
pages = {415--423},
title = {{Entangled quantum systems and the Schmidt decomposition}},
volume = {63},
year = {1995}
}

@article{Aravind1996,
author = {Aravind, P. K.},
journal = {Am. J. Phys.},
pages = {1143--1150},
title = {{Geometry of the Schmidt decomposition and Hardy’s theorem}},
volume = {64},
year = {1996}
}

@article{Zyczkowski1999,
author = {\.{Z}yczkowski, K.},
journal = {Phys. Rev. A},
pages = {3496--3507},
title = {{Volume of the set of separable states. II}},
volume = {60},
year = {1999}
}

@article{Werner1989,
author = {Werner, R. F.},
journal = {Phys. Rev. A},
pages = {4277--4281},
title = {{Quantum states with Einstein-Podolsky-Rosen correlations admitting a hidden-variable model}},
volume = {40},
year = {1989}
}

@article{Castillo2024,
author = {Castillo, A. and Pineda, C. and Navarrete, E. S. and Davalos, D.},
journal = {Phys. Rev. A},
pages = {032204},
title = {{Coarse-grained dynamics in quantum many-body systems using the maximum entropy principle}},
volume = {112},
year = {2025}
}

@article{Busch1993,
author = {Busch, P. and Quadt, R.},
journal = {Int. J. Theor. Phys.},
pages = {2261--2269},
title = {{Concepts of coarse graining in quantum mechanics}},
volume = {32},
year = {1993}
}

@article{Quadt1994,
  author    = {Ralf Quadt and Paul Busch},
  title     = {Coarse grain\-ing and the quan\-tum---Classical connection},
  journal   = {Open Systems \& Information Dynamics},
  year      = {1994},
  volume    = {2},
  number    = {2},
  pages     = {129--155},
  doi       = {10.1007/BF02228961},
  url       = {https://doi.org/10.1007/BF02228961},
  issn      = {1573-1324}
}

@article{Duarte2017,
author = {Duarte, C. and Carvalho, G. D. and Bernandes, N. K. and De melo, F.},
journal = {Phys. Rev. A},
pages = {032113},
title = {{Emerging dynamics arising from coarse-grained quantum systems}},
volume = {96},
year = {2017}
}

@article{Eisert2020,
author = {Eisert, J. and Hangleiter, D. and Walk, N. and Roth, I. and Markham, D. and Parekh, R. Chabaud, U and Kashefi, E.},
journal = {Nature Rev. Phys.},
pages = {382--390},
title = {{Quantum certification and benchmarking}},
volume = {2},
year = {2020}
}

@article{Altman2021,
author = {Altman, E. and Brown, K. R. and Carleo, G. and Carr, L. D. and Demler, E. and Chin, C. and DeMarco, B.},
journal = {PRX Quantum},
pages = {017003},
title = {{Quantum simulators: Architectures and opportunities}},
volume = {2},
year = {2021}
}

@article{Pelucchi2022,
author = {Pelucchi, E. and Fagas, G. and Aharonovich, I. and Englund, D. and Figueroa, E. and Gong, Q. and Hannes, H. and Liu, J. and Lu, C. -Y. and Matsuda, N.},
journal = {Nature Rev. Phys.},
pages = {194--208},
title = {{The potential and global outlook of integrated photonics for quantum technologies}},
volume = {4},
year = {2022}
}

@article{Rosset2012,
author = {Rosset, D. and {Ferretti{-}Sch{\"o}bitz}, R. and Bancal, J. -D. and Gisin, N. and Liang, Y. -C.},
journal = {Phys. Rev. A},
pages = {062325},
title = {{Imperfect measurement settings: Implications for quantum state tomography and entanglement witnesses}},
volume = {86},
year = {2012}
}

@article{Naikoo2021,
  title     = {Projective measurements under qubit quantum channels},
  author    = {Javid Naikoo and Subhashish Banerjee and A. K. Pan and Sibasish Ghosh},
  journal   = {Phys. Rev. A},
  volume    = {104},
  number    = {4},
  pages     = {042608},
  year      = {2021},
}

@article{Wu2011,
  title     = {Non-disturbance Criteria of Quantum Measurements},
  author    = {Wu, Zhaoqi and Zhang, Shifang and Wu, Junde},
  journal   = {Int. J. Theor. Phys.},
  volume    = {50},
  number    = {5},
  pages     = {1325--1333},
  year      = {2011},
}

@article{Zhu2023,
  author    = {Jialiang Zhu and Zhaorong Liu and Zeyang Liao and Shengjun Wu},
  title     = {Learning Informative Latent Representation for Quantum State Tomography},
  journal   = {Phys. Rev. A},
  volume    = {107},
  number    = {3},
  pages     = {032412},
  year      = {2023},
}

@article{Gao2022,
  author    = {Xianxin Gao and Mengmeng Sun and Fengyu Zhang and Kai Xu and Shijie Gu and Xiuhao Deng and Youpeng Zhang and Hui Wang},
  title     = {Deep Learning-Based Quantum State Tomography With Imperfect Measurement},
  journal   = {International Journal of Theoretical Physics},
  volume    = {61},
  number    = {5},
  pages     = {134},
  year      = {2022},
}

@article{Lange2023,
  author    = {Hannah Lange and Matja{\v{z}} Kebri{\v{z}} and, Maximilian Buser and Ulrich Schollw{\"o}ck and Fabian Grusdt and Annabelle Bohrdt},
  title     = {Adaptive Quantum State Tomography with Active Learning},
  journal   = {Quantum},
  volume    = {7},
  pages     = {1129--1152},
  year      = {2023},
}

@book{Heisenberg1958,
author = {Heisenberg, Werner},
title = {Physics and Philosophy: The Revolution in Modern Science},
publisher = {Harper \& Brothers Publishers},
address = {New York},
year = {1958},
series = {World Perspectives},
volume = {19}
}

@book{Heisenberg1971,
author = {Heisenberg, Werner},
title = {Physics and Beyond: Encounters and Conversations},
translator = {Pomerans, Arnold J.},
publisher = {Harper \& Row, Publishers},
address = {New York, Evanston, and London},
year = {1971},
series = {World Perspectives},
volume = {42}
}

@article{Ibrahim2015,
author = {Saideh, I. and Ribeiro, A. D. and Ferrini, G. and Coudreau, T. and Milman, P. and Keller, A.},
journal = {Phys. Rev. A},
pages = {052334},
title = {{General dichotomization procedure to provide qudit entanglement criteria}},
volume = {92},
year = {2015}
}

@article{Yang2021,
  author  = {J. Yang and L. Liu and J. Mongkolkiattichai and P. Schauss},
  titlex   = {Site-resolved imaging of ultracold fermions in a triangular-lattice quantum gas microscope},
title = {Site-re\-solved im\-ag\-ing of ul\-tra\-cold fer\-mi\-ons in a tri\-an\-gu\-lar-lat\-tice quan\-tum gas mi\-cro\-scope},
  journal = {PRX Quantum},
  volume  = {2},
  number  = {2},
  pages   = {020344},
  year    = {2021},
}

@article{Kwon2022,
author = {Kiryang Kwon and Kyungtae Kim and Junhyeok Hur and SeungJung Huh and Jae-yoon Choi},
title = {Site-resolved imaging of a bosonic {M}ott insulator of $^7$Li atoms},
journal = {Phys. Rev. A},
volume = {105},
number = {3},
pages = {033323},
year = {2022},
}

@article{Sherson2010,
author = {Jacob F. Sherson and Christof Weitenberg and Manuel Endres and Marc Cheneau and Immanuel Bloch and Stefan Kuhr},
title = {Sin\-gle-atom-re\-solved fluorescence im\-ag\-ing of an atomic {M}ott insulator},
xtitle = {xxxx},
journal = {Nature},
year = {2010},
volume = {467},
pages = {68--73},
number = {7311}
}
\end{document}